\documentclass[useAMS,usenatbib,usegraphicx]{mn2e}

\usepackage{epsfig,color,float,threeparttable}

\title[A hot Jupiter found in the POTS]{A hot Jupiter transiting a
  mid-K dwarf found in the pre-OmegaCam Transit Survey\thanks{Based on
    observations obtained at Paranal and La Silla Observatories in
    European Southern Observatory (ESO) programmes 076.A-9014(A),
    077.A-9007(A), 077.A-9007(B), 077.C-0659(A), 077.C-0780(A),
    078.A-9057(A), 079.A-9003(A), 083.A-9001(B), 084.A-9002(D),
    383.C-0821(A), 385.C-0817(A) and 086.C-0600(B)}}

\author[Koppenhoefer et al.]  {J.~Koppenhoefer$^{1,2}$\thanks{E-mail:
    koppenh@mpe.mpg.de}, R.~P.~Saglia$^{2,1}$, L.~Fossati$^{3}$,
  Y.~Lyubchik$^{4}$, M.~Mugrauer$^{5}$,\and R.~Bender$^{2,1}$,
  C.-H.~Lee$^{1,2}$, A.~Riffeser$^{1}$, P.~Afonso$^{2,6}$,
  J.~Greiner$^{2}$, Th.~Henning$^{7}$,\and R.~Neuh\"auser$^{5}$,
  I.~A.~G.~Snellen$^{8}$, Y.~Pavlenko$^{4}$, M.~Verdugo$^{2,9}$ 
  and N.~Vogt$^{10}$\\\\
  $^{1}$University Observatory Munich, Scheinerstrasse 1, D-81679
  M\"unchen, Germany\\
  $^{2}$Max Planck Institute for Extraterrestrial Physics,
  Giessenbachstrasse, D-85748 Garching, Germany\\
  $^{3}$Argelander-Institut f\"ur Astronomie der Universit\"at
  Bonn, Auf dem H\"ugel 71, D-53121 Bonn, Germany\\
  $^{4}$Main Astronomical Observatory, ZAbolotnoho 27, Kyiv-127
  03680, Ukraine\\
  $^{5}$Astrophysikalisches Institut und Universit\"ats-Sternwarte,
  Schillerg\"asschen 2, D-07745 Jena, Germany\\
  $^{6}$Physics and Astronomy Department, American River College, 4700
  College Oak Drive, Sacramento, CA 95841, USA\\
  $^{7}$Max Planck Institute for Astronomy, K\"onigstuhl 17, D-69117
  Heidelberg, Germany\\
  $^{8}$Sterrewacht Leiden, PO Box 9513, NL-2300 RA Leiden, The
  Netherlands\\
  $^{9}$Department of Astronomy, University of Vienna, 
  T\"urkenschanzstra\ss e 17, 1180 Vienna, Austria\\
  $^{10}$Departamento de F\'isica y Astronom\'ia, Universidad de
  Valparaiso, Avda. Gran Breta\~na 1111, Valparaiso, Chile}

\begin{document}

\def\Teff{$T_{\mathrm{eff}}$}
\def\logg{\ensuremath{\log g}}
\def\loggf{\ensuremath{\log gf}}
\def\vmic{$\upsilon_{\mathrm{mic}}$}
\def\vsini{\ensuremath{{\upsilon}\sin i}}
\def\kms{$\mathrm{km\,s}^{-1}$}
\def\ms{$\mathrm{m\,s}^{-1}$}

\date{Accepted 2013 August 8. Received 2013 August 8; in original form
  2013 April 16}

\pagerange{\pageref{firstpage}--\pageref{lastpage}} \pubyear{2013}

\maketitle

\label{firstpage}

\begin{abstract}
  We describe the pre-OmegaTranS project, a deep survey for transiting
  extra-solar planets in the Carina region of the Galactic disc. In
  2006--2008, we observed a single dense stellar field with a very
  high cadence of $\sim$2\,min using the European Southern Observatory
  Wide Field Imager at the La Silla Observatory.\\ Using the
  Astronomical Wide-field Imaging System for Europe environment and
  the Munich Difference Imaging Analysis pipeline, a module that has
  been developed for this project, we created the light curves of
  16000 stars with more than 4000 data points which we searched for
  periodic transit signals using a box-fitting least-squares detection
  algorithm. All light curves are publicly available. In the course of
  the pre-OmegaTranS project, we identified two planet candidates --
  POTS-1b and POTS-C2b -- which we present in this work.\\ With
  extensive follow-up observations we were able to confirm one of
  them, POTS-1b, a hot Jupiter transiting a mid-K dwarf. The planet
  has a mass of 2.31$\pm$0.77\,M$_{\rm{Jup}}$ and a radius of
  0.94$\pm$0.04\,R$_{\rm{Jup}}$ and a period of $P$\,=\,3.16\,d. The
  host star POTS-1 has a radius of 0.59$\pm$0.02\,R$_{\odot}$ and a
  mass of 0.70$\pm$0.05\,M$_{\rm{\odot}}$. Due to its low apparent
  brightness of $I$\,=\,16.1\,mag the follow-up and confirmation of
  POTS-1b was particularly challenging and costly.\\
\end{abstract}


\section{Introduction}
\label{sec.intro}
The field of transiting extra-solar planet detection has changed a lot
in the last decade. After the first transit observations of the planet
HD\,209458b \citep{2000ApJ...529L..45C} that had been detected using
the radial velocity (RV) method, the OGLE-III deep survey
\citep{2004AcA....54..313U} was able to deliver the first detections
of transiting planet candidates which were successfully confirmed with
RV follow-up observations. The characteristics of this project allowed
the detection of planets around faint stars which made any follow-up
study difficult and expensive in terms of observing
time. Nevertheless, the OGLE-III survey was the first to provide an
estimate on the fraction of stars hosting a previously undetected
population of very hot Jupiters with periods below 3\,d
\citep{2007A&A...475..729F,2008ApJ...686.1302B}. At the same time,
several wide-angle surveys were initiated which target brighter stars
using small-aperture optics such as TRES \citep{2007ApJ...663L..37O},
XO \citep{2005PASP..117..783M}, HAT \citep{2004PASP..116..266B} and
WASP \citep{2006PASP..118.1407P}.  The number of known transiting
planets was rising very quickly with each of them being an interesting
target for follow-up studies such as the measurement of the
spin--orbit alignment \citep[e.g.][]{2009ApJ...703L..99W}, thermal
emission \citep[e.g.][]{2005ApJ...626..523C}, transit timing
variations \citep[e.g.][]{2010MNRAS.407.2625M,2011PASJ...63..287F} or
transmission spectroscopy studies
\citep[e.g.][]{2012ApJ...747...35B}.\\ The SWEEPS survey
\citep{2006Natur.443..534S} utilized the {\it Hubble Space Telescope}
to search for short-period transiting planets in the Galactic bulge
and found two planets and 14 additional candidates which are too faint
($V$\,$>$\,19\,mag) to be followed up and confirmed with current
instrumentation. The launch of the {\it CoRoT} and {\it Kepler}
satellites \citep{2009A&A...506..411A,2010Sci...327..977B} marked a
new age of transit detection. An unprecedented precision of typically
10$^{-5}$ allows those missions to study the population of extra-solar
planets down to Earth radii with great statistics. Up to now, 2312
planet candidates have been published by the Kepler team
\citep{2013ApJS..204...24B} and even if the confirmation by RV
measurements will be difficult and will take time, the expectation is
that many of the detected objects are indeed planets
\citep{2011ApJ...738..170M} although there are indications that the
false positive rate varies significantly with planetary and stellar
properties \citep{2012MNRAS.426..342C}.\\ Another comparably young
branch of projects is formed by ground-based surveys dedicated to the
detection of planets around M dwarfs. The MEarth project
\citep{2009IAUS..253...37I} is monitoring bright M dwarfs with a
network of small telescopes. The WFCAM Transit Survey
\citep{2013MNRAS.tmp.1446K} is a deep J-band survey for planets around
cool stars. M dwarfs have the advantage that their habitable zone is
much closer compared to earlier spectral types making the detection of
planets in this distance easier. Two large-area deep surveys, the
Palomar Transient Factory \citep{2011ASPC..448.1367L} and Pan-Planets
\citep{2009A&A...494..707K}, are aiming at the detection of Jupiter-
to Neptune-sized planets by observing of the order of 100000 M
dwarfs.\\ In this work, we present the final results of the
pre-OmegaTranS (POTS) project which is a pilot study for a larger
scale transit survey that was planned to be conducted with the ESO
Very Large Telescope (VLT) Survey Telescope (VST). In Section
\ref{sec.POTS}, we give an overview of the POTS including a
description of the data reduction and light curve analysis. We present
two planet candidates that we found in the survey. Section
\ref{sec.followup} summarizes several follow-up studies we conducted
in order to confirm the planetary nature of our best candidate --
POTS-1b. Using both photometric and spectroscopic observations, we
were able to derive the mass and radius of the planet. Making use of
high-resolution imaging with adaptive optics (AO) as well as a
detailed analysis of the light curves enabled us to exclude all
possible blend scenarios. In section \ref{sec.POTS-C2}, we present our
second best candidate, POTS-C2 and in Section \ref{sec.conclusions} we
discuss the results and draw our conclusions.\\
\section{The POTS Project}
\label{sec.POTS}
In 2004, a consortium of astronomers from INAF Capodimonte (Italy),
Sterrewacht Leiden (Netherlands) and MPE Garching (Germany) designed
the OmegaCam Transit Survey (OmegaTranS). A total of 26 nights of
guaranteed time observations with OmegaCam \citep{2002Msngr.110...15K}
at the VST \citep{2002SPIE.4836...43C} were granted to this project by
the three institutes. Scaling from existing surveys, such as
OmegaTranS, was expected to deliver 10--15 new detections per year
with the main power being the large 1 square degree field of view
(FoV) of the OmegaCam detector.\\ Due to delays in the construction
and commissioning of the telescope, the start of the project has been
delayed and ultimately cancelled. Instead, we conducted the POTS using
the ESO Wide Field Imager (WFI) mounted on the 2.2\,m telescope at the
La Silla observatory \citep{1999Msngr..95...15B}.\\ In Section
\ref{subsec.characterization}, we present a characterization of the
POTS target population. We obtained absolute magnitudes in the $U$,
$B$, $V$, $R$ and $I$ filters and colours for each star in our target
field. In Section \ref{subsec.observations}, we give an overview of
the POTS data collected in three observational seasons 2006--2008. The
creation of the light curves using the difference imaging technique
and the light-curve analysis and candidate selection are presented in
Section \ref{subsec.lc_analysis}.\\
\subsection{Characterization of the POTS target population}
\label{subsec.characterization}
\begin{figure}
  \centering
  \includegraphics[width=0.45\textwidth]{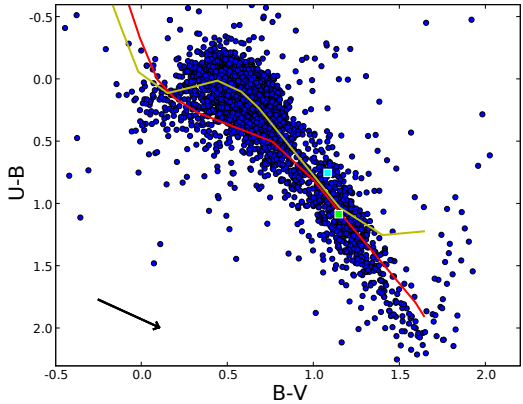}
  \caption{Extinction corrected $UBV$ colour--colour diagram of the
    3000 brightest stars in the OTSF-1a field. The yellow and red
    lines show the position of the main-sequence and luminosity class
    III giants \citep[according to][]{1998gaas.book.....B}. The green
    and cyan squares show the positions of the planet POTS-1 and the
    candidate POTS-C2 which are presented in this work (see the
    text). The black arrow shows the extinction correction that has
    been applied.}
  \label{fig.UBV} 
\end{figure}

In the course of the POTS, we observed one single WFI field (OTSF-1a)
which is the north-west corner of the previously selected best
OmegaTranS field \mbox{OTSF-1} \citep{2007ASPC..366...78B}. The image
centre is RA\,=\,13$^h$35$^m$41$\fs$6 and
Dec.\,=\,--$66^{\circ}$42'21'' and the field dimensions are
34\,arcmin\,$\times$\,33\,arcmin.\\ In order to determine absolute
magnitudes and colours for each star in our target field, we performed
a photometric calibration. In 2006 March, we obtained observations of
\mbox{OTSF-1a} in the $U$, $B$, $V$, $R$ and $I$ bands (filter \#877,
\#878, \#843, \#844 and \#879) as well as a set of standard star
observations at several different airmasses.\\ We transformed the
measured magnitude of each star into a calibrated magnitude in the
Johnson--Cousins magnitude system using a filter-dependent extinction
and a colour term. We assume the colour coefficients to be fixed
properties of the WFI filters which are constant over time and
therefore adapted the standard values obtained from the ESO/WFI
instrument
page.\footnote{http://www.eso.org/sci/facilities/lasilla/instruments/wfi}
The extinction coefficients were measured using multiple observations
of standard star fields \citep{1992AJ....104..340L} which were taken
at various airmasses (except for the $U$-band calibration where too
few standard star field observations were available; we used the
$U$-band extinction coefficient published on the WFI web page).\\ We
performed aperture photometry on all standard stars using a large
aperture with 30 pixel diameter ($\hat{=}$ 6\,arcsec) which is much
larger than the maximum full width at half maximum (FWHM) of the point
spread functions (PSFs) of all images ($\sim$2\,arcsec) and compared
our measurements to a reference catalogue that has been exported from
the Astronomical Wide-field Imaging System for Europe (Astro-WISE)
\citep{2007ASPC..376..491V} and which contains measurements from
\citet{1992AJ....104..340L}, \citet{2000PASP..112..925S} and from the
Sloan Digital Sky Survey (SDSS) DR5 \citep{2007ApJS..172..634A}. SDSS
measurements were transformed into the Johnson--Cousins filter system
using the equations given in \citet{2005AJ....130..873J}.\\ We assumed
the extinction coefficient to be independent of CCD number and used
the average of the values we obtained for the individual CCDs. Table
\ref{tab.extcol} lists the measured extinction coefficients for the
$B$, $V$, $R$ and $I$ bands as well as the extinction coefficient of
the $U$ band and all colour coefficients that were taken from the WFI
web page.\\ After the extinction and colour term correction, we
derived a zero-point for each CCD and each filter by fitting a
constant offset to the residuals. In order to check if we had to
correct for zero-points variations over the FoV of one CCD (known as
illumination correction), we analysed the residuals as a function of
$x$- and $y$-position on the CCD and found no trends. We therefore
used a constant zero-point for each CCD.\\ Table \ref{tab.zeropoints}
lists our measured zero-points. The errors of the zero-points were
estimated from the rms of the final residuals.\\
\begin{table*}
  \centering
  \begin{tabular}{l|r|r|r|r|r}\hline
                           & \multicolumn{1}{|c}{$U$} & \multicolumn{1}{|c}{$B$} & \multicolumn{1}{|c}{$V$} & \multicolumn{1}{|c}{$R$} & \multicolumn{1}{|c}{$I$}  \\ \hline
    Extinction coefficient & 0.48 & 0.23 &   0.18 & 0.16 & 0.11 \\
    Colour coefficient     & 0.05 & 0.25 & --0.13 & 0.00 & 0.03 \\\hline
  \end{tabular}
  \caption{Extinction and colour coefficients used in this work.}
  \label{tab.extcol}
\end{table*}
\begin{table*}
  \centering
  \begin{tabular}{c|c|c|c|c|c}\hline
    CCD   & $ZP_U$ (mag)    & $ZP_B$ (mag)    & $ZP_V$ (mag)    & $ZP_R$ (mag)    & $ZP_I$ (mag)    \\ \hline
    ccd50 & 22.32$\pm$0.06  & 24.77$\pm$0.03  & 24.12$\pm$0.04  & 24.46$\pm$0.05  & 23.36$\pm$0.09  \\
    ccd51 & 22.20$\pm$0.11  & 24.88$\pm$0.02  & 24.24$\pm$0.03  & 24.55$\pm$0.03  & 23.48$\pm$0.05  \\
    ccd52 & 22.20$\pm$0.09  & 24.66$\pm$0.04  & 24.01$\pm$0.03  & 24.37$\pm$0.06  & 23.29$\pm$0.07  \\
    ccd53 & 22.18$\pm$0.06  & 24.83$\pm$0.05  & 24.20$\pm$0.03  & 24.55$\pm$0.09  & 23.42$\pm$0.08  \\
    ccd54 & 22.24$\pm$0.12  & 24.80$\pm$0.03  & 24.15$\pm$0.04  & 24.54$\pm$0.09  & 23.41$\pm$0.08  \\
    ccd55 & 22.25$\pm$0.12  & 24.89$\pm$0.05  & 24.24$\pm$0.04  & 24.58$\pm$0.07  & 23.47$\pm$0.09  \\
    ccd56 & 22.14$\pm$0.10  & 24.82$\pm$0.04  & 24.19$\pm$0.04  & 24.51$\pm$0.05  & 23.39$\pm$0.05  \\
    ccd57 & 22.21$\pm$0.11  & 24.66$\pm$0.04  & 24.03$\pm$0.03  & 24.39$\pm$0.06  & 23.31$\pm$0.07  \\\hline
  \end{tabular}
  \caption{Measured U-, B-, V-, R- and I-band zero-points for each
    CCD.}
  \label{tab.zeropoints}
\end{table*}
\noindent In order to derive the $U$-, $B$-, $V$-, $R$- and $I$-band
magnitudes of all stars in our target field, we performed aperture
photometry on each of the OTSF-1a images. Since the field is very
crowded and our target stars are comparably faint, we used a small
aperture of 15\,pixels and applied an aperture correction to the
measured fluxes. The aperture correction was determined by comparing
the 15 and 30\,pixel aperture fluxes of bright stars. For each filter,
three observations of OTSF-1a were taken which gives us up to three
independent measurements of each star. We use the median value for
each star.\\ In a final step, we corrected our measurements for
Galactic extinction.  From \citet{1998ApJ...500..525S}, we obtained
the total reddening for extragalactic objects
$E(B-V)$\,=\,0.698. Assuming a standard extinction law, this
translates in \mbox{$E(U-B)$\,=\,0.64\,$\times$\,$E(B-V)$\,=\,0.45}
\citep{1998gaas.book.....B}. Since our target stars are located inside
the Galaxy, the actual reddening should be lower than the
extragalactic reddening (depending on the individual distance of each
star). Fig. \ref{fig.UBV} shows the $UBV$ colour--colour plot of the
OTSF-1a stars after correcting for an average reddening of half the
extragalactic value \citep{1998ApJ...500..525S}. For individual
sources (i.e. our detected transit candidates), we performed a more
detailed analysis and derived individual extinction values and
distances (see Section \ref{subsec.lc_analysis}). Yellow and red lines
indicate the location of main-sequence dwarfs and luminosity class III
giants \citep[according to][]{1998gaas.book.....B}.\\
\subsection{Photometric observing campaigns 2006--2008}
\label{subsec.observations}
\begin{figure}
  \centering
  \includegraphics[width=0.45\textwidth]{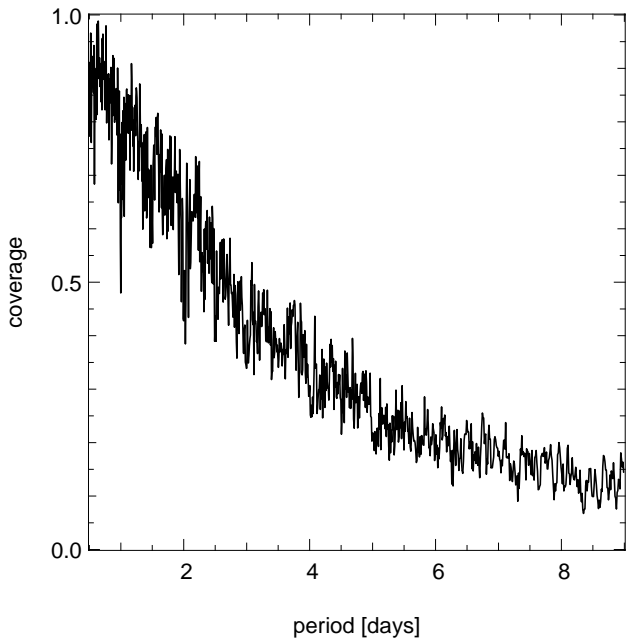}
  \caption{Orbital phase coverage of the POTS light curves. The line
    shows the probability to see at least two transits for a given
    orbital period.}
  \label{fig.coverage} 
\end{figure}

A total of 129\,h of observations were collected in the years
2006--2008. Spread over 34 nights, we obtained 4433 epochs in the
Johnson $R$ band (filter \#844). The exposure time was 25\,s in most
cases. Under very good and very bad observing conditions, we slightly
adjusted the exposure time in order to achieve a stable
signal-to-noise ratio (S/N) or to avoid saturating too many stars. The
average cadence (exposure, readout and file transfer time) is
107\,s. The median seeing is 1.6\,arcsec. 167 images with a seeing
larger than 2.5\,arcsec were not used because of their bad
quality. Fig. \ref{fig.coverage} shows the probability to witness two
or more transits as a function of the orbital period. The survey
sensitivity drops quickly towards longer periods due to the limited
amount of observing time that was available for this pilot
study. \\ In addition to the science images, we obtained calibration
images (i.e. bias and flat-field exposures) for each of the 34
nights.\\
\subsection{Data analysis and light-curve extraction}
\label{subsec.data_analysis}
The basic CCD data reduction steps were done using the Astro-WISE
standard calibration pipeline \citep{2007ASPC..376..491V}. The
processing steps include overscan and bias correction, flat-fielding
and masking of satellite tracks, bad pixels and cosmics. The data
reduction pipeline treats all CCDs as independent detectors. For a
complete description of all tasks, we refer to the Astro-WISE User and
Developer
manual.\footnote{http://www.astro-wise.org/docs/Manual.pdf}\\ All
images of one CCD roughly map the same part of the sky. However, small
shifts of the individual exposures (smaller than 20 pixels) arise from
a limited pointing accuracy and a small dithering.\\ We calculated for
each CCD the absolute astrometric solution for the very best seeing
image by measuring the positions and brightnesses of several hundred
stars with \textsc{sextractor} \citep{1996A&AS..117..393B} and
comparing them to the positions in the USNO-A2.0 catalogue. Using
least-squares methods, a transformation was calculated that corrects
for a shift, a rotation and a third-order polynomial distortion.\\ We
combined the 15 best seeing images of each CCD (typically around
0.6\,arcsec FWHM) to create a reference stack which was then used to
perform a relative astrometric calibration of all single images. In
this way, we achieved a median rms of the astrometric solutions of
40\,mas. Note that a very good overlap is important for the difference
imaging approach (see below). Out of the 4266 single images, 12 had an
rms of the astrometric residuals that was larger than 0.1\,arcsec and
were not used in the following steps. Each image was resampled to a
new grid with a pixel scale of 0.2\,arcsec\,pixel$^{-1}$ using the
program
\textsc{swarp}\footnote{http://www.astromatic.net/software/swarp}
which uses a LANCZOS3 interpolation algorithm.\\ We applied the Munich
Difference Imaging Analysis (\textsc{mdia}) package
\citep{2013ExA....35..329K} to the resampled images in order to create
light curves using the difference imaging method
\citep{1996AJ....112.2872T,Alard}. The technique has become the most
successful method used for the creation of high-precision light curves
in crowded fields such as the Milky Way bulge
\citep{2008AcA....58...69U} or the Andromeda Galaxy
\citep{2008ApJ...684.1093R}. The \textsc{mdia} package is based on the
implementation presented in \citet{2002A&A...381.1095G}.\\ The method
uses a reference image which is degraded by convolution in order to
match the seeing of each single image in the data set.  Subtracting
the convolved reference image from a single image, one gets a
so-called difference image with all constant sources being removed and
variable sources appearing as positive or negative PSF-shaped
residuals.\\ In the difference imaging process, we adopted the
standard parametrization of the kernel base functions \citep{Alard}
and a kernel size of 41\,$\times$\,41\,pixels, together with a
third-order polynomial to account for background differences. The 60
free parameters were determined via $\chi^2$-minimization. We used
almost all pixels to determine the optimal kernel and background
coefficients. Only pixels that belong to variable objects were not
taken into account since these would have destroyed the normalization
of the kernel. As we did not know a priori which pixel belong to
variable objects, we first created a subset of difference images
without masking any pixel, identified variable objects and masked them
in the second run when we created all difference images. In order to
account for small variations of the PSF over the FoV of the detector,
we split each image into 4\,$\times$\,8 subfields and determine a
kernel in each of the subfields independently.\\ To construct the
light curves of each object, we combine the differential fluxes
measured in the difference images with the constant flux measured in
the reference image. The flux measurements in the reference image were
done with \textsc{usmphot}\footnote{\textsc{usmphot} is part of the
  \textsc{mupipe} data analysis package:
  http://www.usm.lmu.de/$\sim$arri/mupipe} which is an iterative
PSF-fitting program that is very similar to \textsc{daophot}
\citep{1987PASP...99..191S}.\\ The photometry on the difference image
was done also with PSF photometry. We constructed the PSF from the
normalized convolved reference image using the same isolated stars we
used to measure the fluxes in the reference image. Note that in this
way we automatically obtained the differential fluxes in the same
units as in the reference image which results in correct amplitudes
(i.e. transit depths).\\ In a final step, we normalized the fluxes of
all light curves \mbox{to 1} (i.e. divided by the median flux) and
applied a barycentric time correction using the formulae of
\citet{1982QB51.3.E43M43..} as implemented in the \textsc{skycalc}
program by
Thorstensen.\footnote{http://www.dartmouth.edu/$\sim$physics/faculty/thorstensen.html}\\
\subsection{Light-curve analysis and candidate selection}
\label{subsec.lc_analysis}
\begin{figure}
\centering
\includegraphics[width=0.45\textwidth]{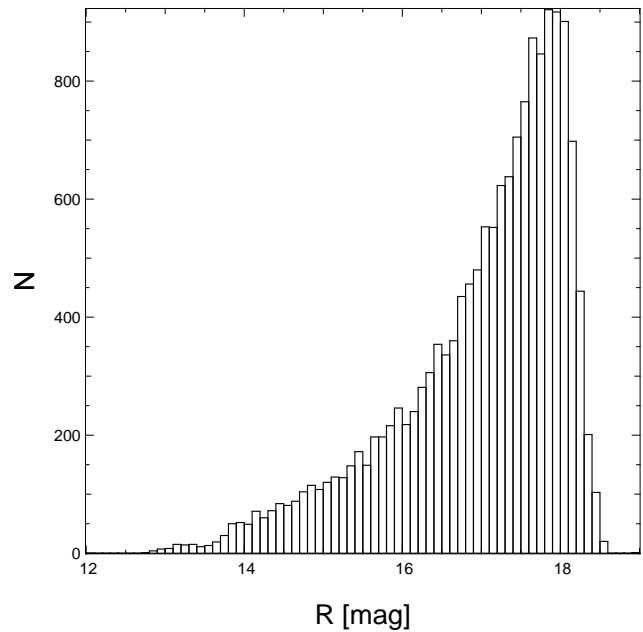}
\caption{Magnitude histogram of all selected sources in the POTS.}
\label{fig.mag_hist}
\end{figure}

\begin{figure}
\centering
\includegraphics[width=0.45\textwidth]{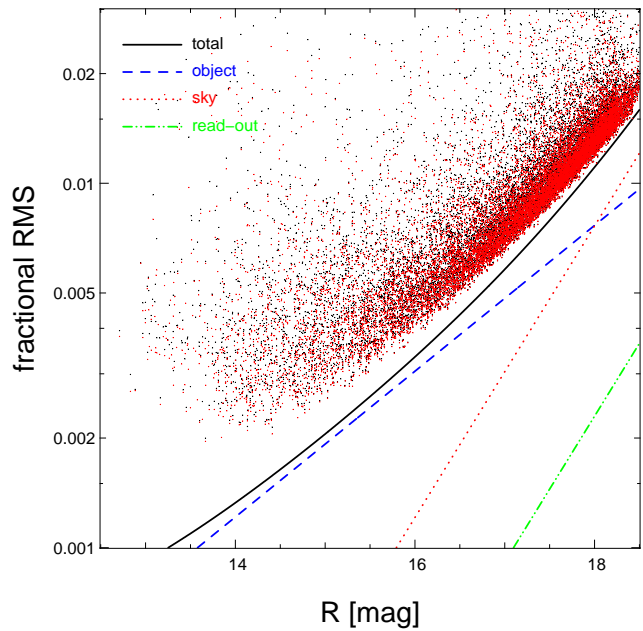}
\caption{rms before (black points) and after (red points) application
  of the $sysrem$ algorithm. The black line shows the theoretical rms
  for our survey parameters. The blue dashed and the red dotted lines
  show the contribution of the photon noise from object and sky
  background, respectively, and the green dash--dotted line shows the
  contribution of the readout noise.}
\label{fig.rms_sysrem}
\end{figure}

\begin{figure}
  \centering
  \begin{minipage}[b]{0.49\linewidth}
    \fbox{\includegraphics[width=0.95\textwidth]{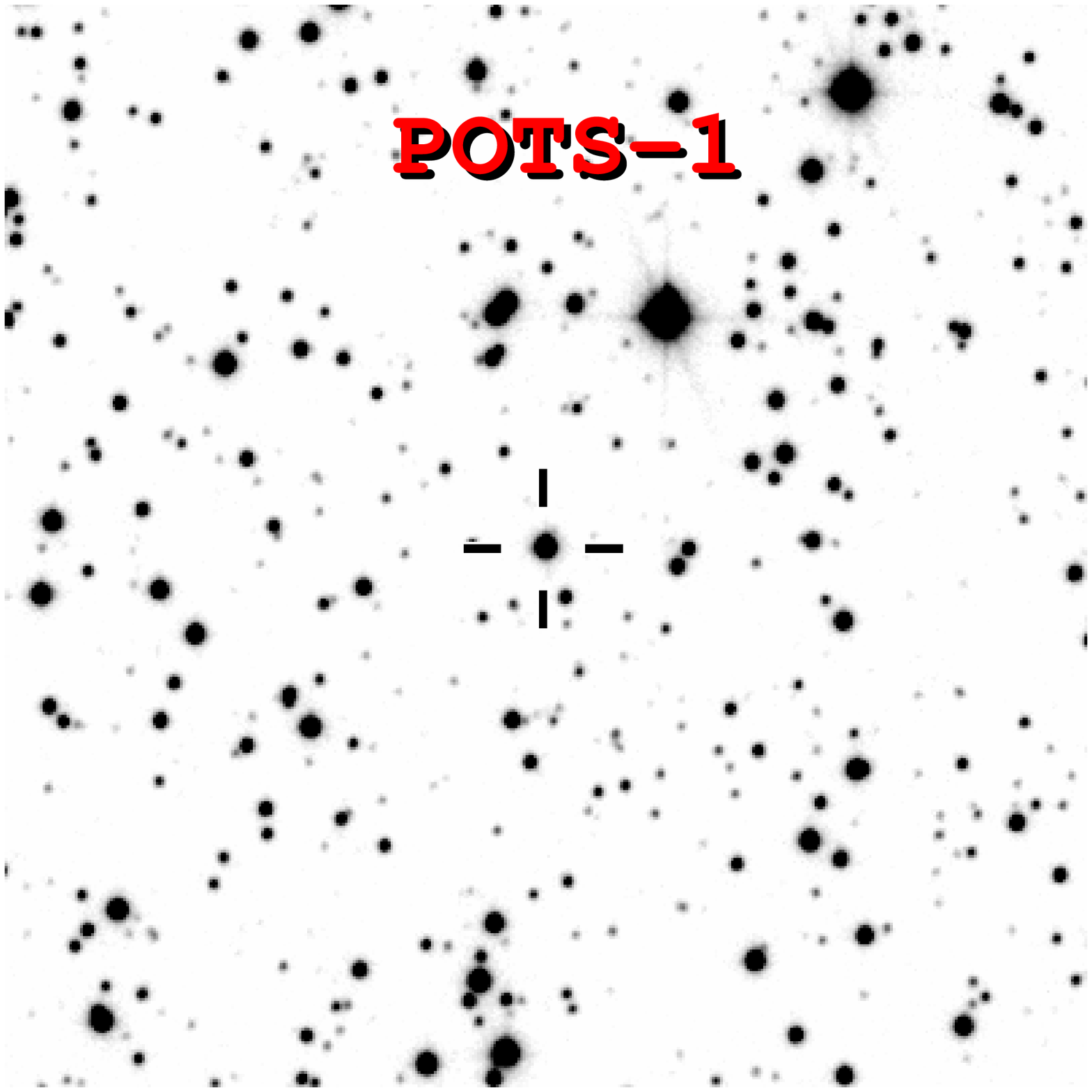}}
  \end{minipage}
  \begin{minipage}[b]{0.49\linewidth}
    \fbox{\includegraphics[width=0.95\textwidth]{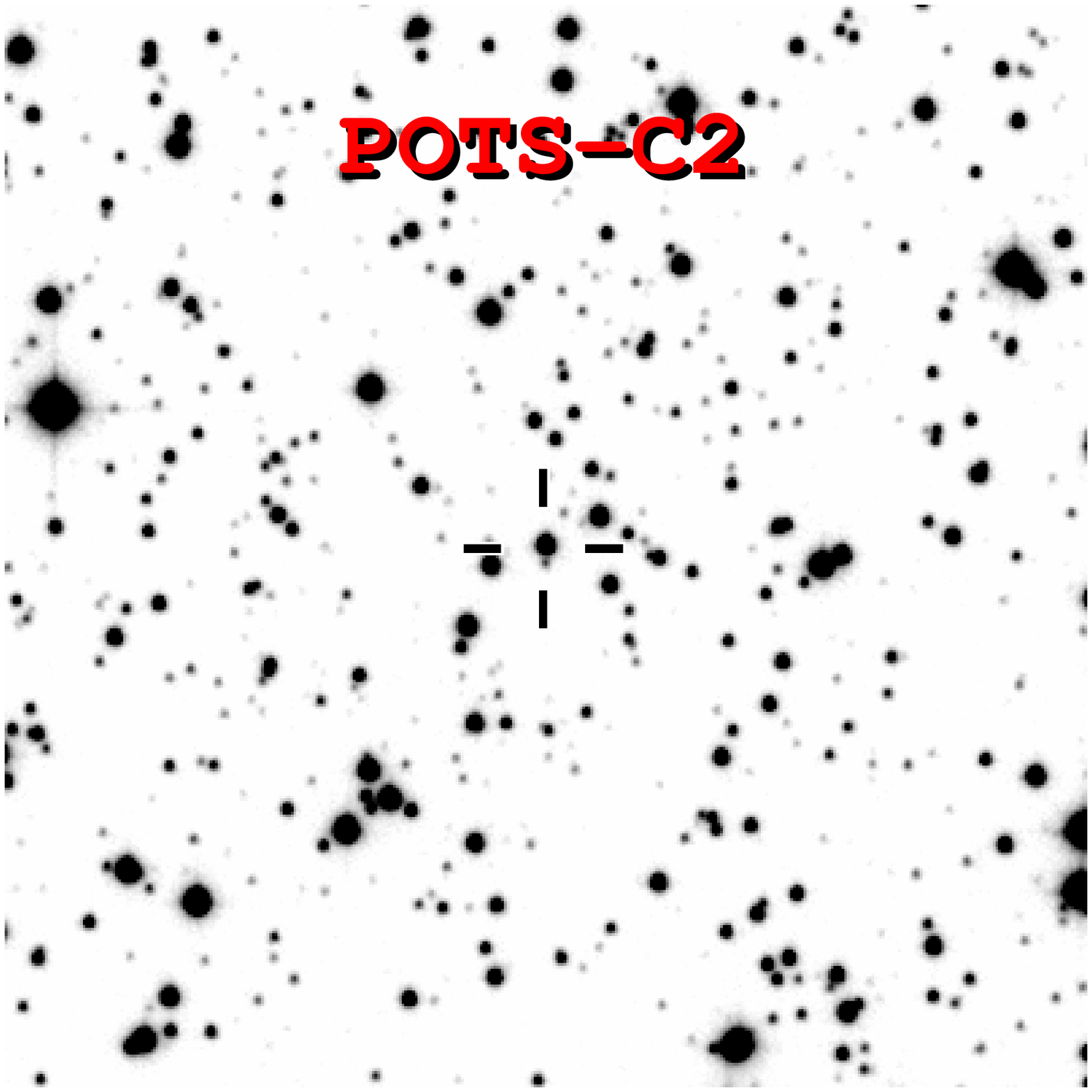}} 
  \end{minipage}
  \caption{Finding chart of POTS-1 and POTS-C2. The FoV is
    \mbox{1.5\,$\times$\,1.5\,arcmin}. North is up and east is left.}
    \label{fig.finding_charts} 
\end{figure}

For each CCD, we extracted the light curves of the 2000 brightest
sources. Fig. \ref{fig.mag_hist} shows the magnitude distribution of
the selected objects. In order to correct for red noise, we applied
the $sysrem$ algorithm \citep{2005MNRAS.356.1466T} which has turned
out to be very successful in reducing systematic effects and which is
used in a large number of transit surveys
\citep[e.g.][]{2006MNRAS.373..231P,2007A&A...476.1357S}.\\ $Sysrem$
works most efficiently if all light curves of variable stars are
removed from the sample. Therefore, we fitted a constant baseline to
each light curve and calculated the reduced $\chi^2$ of the fit. We
applied the $sysrem$ algorithm to all light curves with
$\chi^2$\,$\le$\,1.5 (70 per cent of all light curves) and subtracted
four systematic effects.\\ Fig. \ref{fig.rms_sysrem} shows the rms of
the light curves before (black points) and after the $sysrem$
correction (red points). The black line shows the theoretical rms for
25\,s exposure time, airmass 1.4, sky brightness of 20.3\,mag
arcsec$^{-2}$ and 1.5\,arcsec seeing and 0.0005\,mag scintillation
noise [estimated using the formula of
  \citet{1967AJ.....72..747Y}]. The other lines show the contributions
of the different noise components. Note that the magnitude-independent
scintillation noise is outside the limits of the figure. The $sysrem$
algorithm reduced the rms of the light curves by only a small
amount. We therefore conclude that systematic effects in our data set
are not as prominent as is the case for other surveys \citep[see
  e.g.][]{2006MNRAS.373..231P,2007A&A...476.1357S}. At the bright end,
we reached a photometric precision of $\sim$2--3\,mmag which is close
to the theoretical precision.\\
\\
\\
In order to find transiting planet candidates, we applied the
box-fitting least-squares (BLS) algorithm proposed by
\citet{2002A&A...391..369K} to all $sysrem$-corrected light curves. We
tested 4001 different periods equally spaced in 1/$p$ between 0.9 and
9.1\,d. We used 1000 bins in the folded light curves. Note that the
choice of only 4001 different periods could be questioned since the
survey data span a period of 572\,d which would correspond to 63 or
636 revolutions for the longest and shortest periods, respectively. In
the worst case (i.e. when the true period is exactly between two
tested periods), the uncertainty adds up to a phase shift between the
first and the last observations of 0.07 or 0.04 phase units,
respectively. This is of the same order of the typical fractional
transit duration (0.01 and 0.1 phase units for the longest and
shortest periods) and our choice of 4001 test periods in principle
could have limited our survey sensitivity. However, 95 per cent of the
data were taken within a period of 241\,d (2006 July to 2007 May) and
therefore the impact on the survey sensitivity is not very
strong.\\ We determined the survey sensitivity using Monte Carlo
simulations \citep{2009PhDT.......287K}. The overall efficiency of 23
per cent was relatively low mainly due to the limited amount of total
observing time.\\ For each light curve we determined the best-fitting
period $P$, epoch $t_0$, transit depth \mbox{$\Delta F$/$F$} and
fractional transit length $\tau$.  We also determine the number of
transits, number of data points during a transit, the S/N of the light
curve and the SDE of each detection and calculate the reduced $\chi^2$
of the box fit.\\ As an additional very useful parameter, we measured
the variations that are in the out-of-transit part of the light
curve. After masking the detected signal, we run the BLS algorithm
again on the remaining data points and compare the S/N found in the
masked light curve, S/N$_{\textrm{removed}}$ hereafter, to the S/N
found in the unmasked light curve. In the case of variable stars, the
difference between the two values (S/N\,-\,S/N$_{\textrm{removed}}$)
is expected to be low, whereas for a transiting planet the difference
should be high.\\ In order to identify all interesting transiting
planet candidates and to reject variable stars and other false
positives, we applied three selection criteria: we require a minimum
of two transits (otherwise the period of the orbit could not be
determined). Furthermore, we require S/N\,$\ge$\,12 and
(S/N\,-\,S/N$_{\textrm{removed}}$) $\ge$ 4.7. The last two selection
criteria were optimized using Monte Carlo simulations. More details
about the optimization procedure can be found in
\citet{2009PhDT.......287K}.\\\\
Among $\sim$200 light curves that passed the detection criteria
presented in the previous section, we identified two transiting planet
candidates. The remaining detections were visually classified as
variable stars or false detections caused by systematic outliers. Due
to limited observing time in combination with the faintness of the
candidates we decided to follow up only the best candidate,
i.e. POTS-1. The quality of the candidate POTS-C2 is also good;
however, we decided to concentrate on POTS-1 because it is brighter
(0.8\,mag in $I$) and slightly redder.\\ We present our intensive
spectroscopic and photometric follow-up observations of POTS-1 in
Section \ref{sec.followup} which lead to a confirmation of the
planetary nature of POTS-1b. In Section \ref{sec.POTS-C2}, we present
the candidate POTS-C2 which we propose for follow-up in the
future. Fig. \ref{fig.finding_charts} shows a
1.5\,$\times$\,1.5\,arcmin finding chart for each of the two
objects.\\
\section{Follow-up and confirmation of POTS-1}
\label{sec.followup}
In the last years we performed several complementary follow up studies
which allowed us to confirm one of our candidates, POTS-1b, as a
planet. In section Section \ref{subsec.spectroscopic}, we describe the
spectroscopic observations we used to determine the atmospheric
parameters of the host star as well as the mass of POTS-1b. In Section
\ref{subsec.photometric}, we present the combined analysis of the POTS
light curve together with targeted multiband observations of three
additional transits. With high angular resolution adaptive imaging we
excluded contaminating sources in the vicinity of POTS-1, as described
in Section \ref{subsec.imaging}. Finally, in Section \ref{L9.blend} we
describe how we ruled out possible blend scenarios.\\
\subsection{Spectroscopic observations with UVES}
\label{subsec.spectroscopic}
\subsubsection{Observations and reduction of the spectra}
\label{subsec.spectroscopic1}
We observed POTS-1 with the UV--Visual Echelle Spectrograph
\citep[UVES;][]{2000SPIE.4008..534D}, mounted at the Nasmyth B focus
of UT2 of ESO's VLT at Paranal, Chile. The aim of these observations
was to estimate the spectroscopic parameters of the host star, and to
determine the RV variations which then allowed us to estimate the mass
of the transiting object.\\ In 2009, we collected a total of 10
observations in fibre mode, with UVES connected to the FLAMES facility
\citep{2002Msngr.110....1P}. We used a setup with a central wavelength
of \mbox{580 nm}, resulting in a wavelength coverage of
4785--6817$\AA$ over two CCDs, at a resolving power of
R\,=\,47000. Due to the faintness of the target, we had to use long
exposure times of 46\,min. Three observations turned out to have a
high background contamination due to stray light within the
detector. We were not able to use those data but we got compensation
in the form of a re-execution \mbox{in 2010}.\\ In 2010, we collected
seven observations with UVES in slit mode. The exposure times were
48\,min. We chose a slit width of 1\,arcsec which translates into a
resolving power of R\,=\,40000. We used the standard setup with a
central wavelength of \mbox{520\,nm}, resulting in a wavelength
coverage of 4144$-$6214$\AA$.\\ Both UVES data sets were reduced using
the \textsc{midas}-based pipelines provided by ESO, which result in
fully reduced, wavelength-calibrated spectra. Unfortunately, the
pipeline for fibre-mode observations failed to extract the spectra of
four of our observations.\\ Table \ref{tab.RV_epochs} gives an
overview of the spectroscopic observations that are used in this
work. The S/N per resolution element, in the central part of the red
orders, varied between 10 and 30 for the different epochs.\\
\subsubsection{Spectroscopic analysis of the host star}
\label{subsec.spectroscopic2}
\begin{figure}
  \begin{center}
    \includegraphics[width=0.45\textwidth]{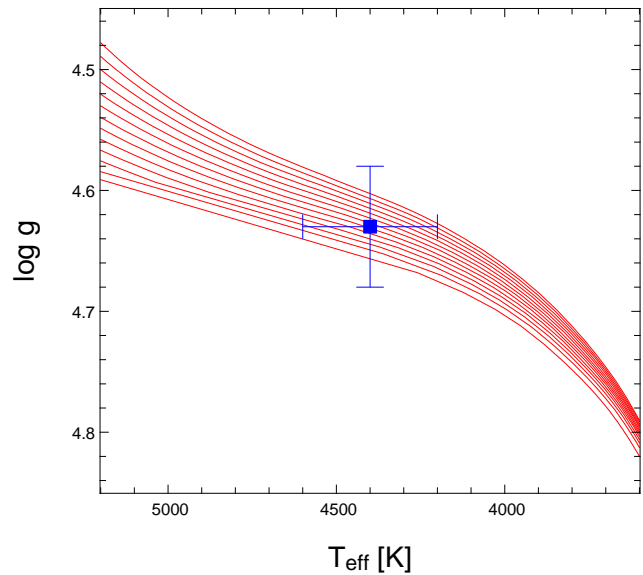}
    \caption{log\,$g$\,--\,\Teff\ relation for low-mass stars showing
      isochrones of 1 to 13\,Gyr for solar metallicity
      \citep{2008ApJS..178...89D}. Assuming that POTS-1 is on the main
      sequence, we derive a \logg\ of 4.63$\pm$0.05 using the estimate
      for \Teff\ derived from SED fitting.}
    \label{isochrones}
  \end{center}
\end{figure}

\begin{figure}
  \begin{center}
    \includegraphics[width=0.45\textwidth]{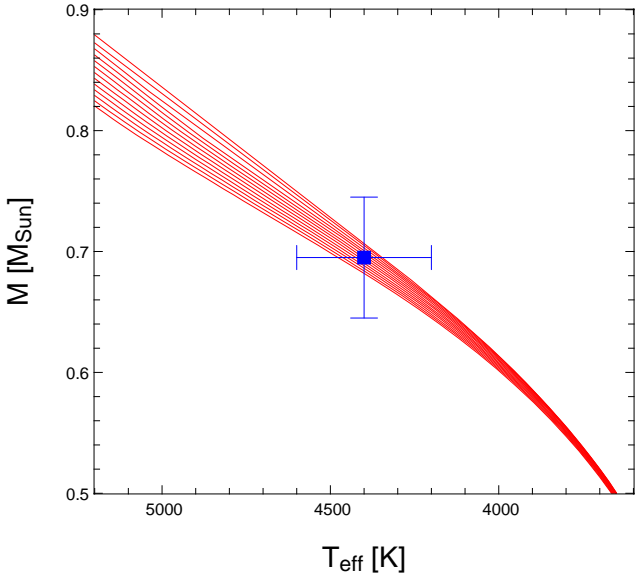}
    \caption{Mass\,--\,\Teff\ relation for low-mass stars. The lines
      show isochrones ranging from 1 to 13\,Gyr for solar metallicity
      \citep{2008ApJS..178...89D}. Assuming that POTS-1 is on the main
      sequence, we derive a mass of 0.695$\pm$0.050\,M$_{\odot}$ using
      the estimate for \Teff\ derived from SED fitting.}
    \label{mass}
  \end{center}
\end{figure}

\begin{figure}
  \begin{center}
    \includegraphics[width=0.45\textwidth]{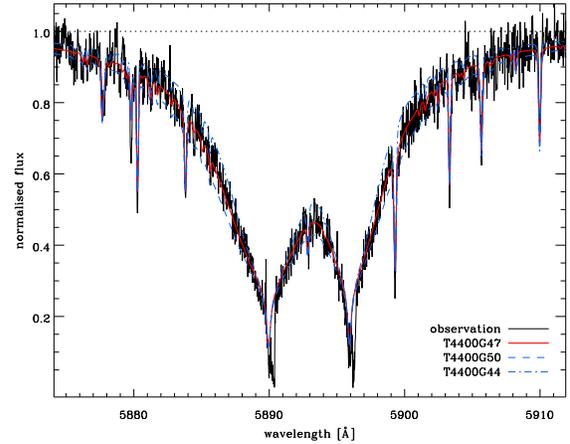}
    \caption{Comparison between the observed Na\,\textsc{i}\,D lines
      (black solid line) and synthetic profiles calculated with our
      final adopted parameters (red thick line) and with
      \logg\ increased by 0.3\,dex (blue dashed line) and decreased by
      0.3\,dex (blue dash--dotted line). Interstellar absorption is
      visible just redwards of the cores of the Na\,\textsc{i}\,D
      lines.}
    \label{fig:sodium}
  \end{center}
\end{figure}

We estimated the atmospheric parameters and chemical abundances of
POTS-1 analysing simultaneously the spectral energy distribution (SED)
and high-resolution spectra obtained combining the seven epochs taken
in slit mode. The combined spectrum has an S/N of 43.4 per resolution
element in the central parts of the spectral region.\\ For both SED
and spectral analysis we employed the \textsc{marcs} stellar model
atmosphere code from \citet{2008A&A...486..951G}. For all
calculations, local thermodynamical equilibrium and plane-parallel
geometry were assumed, and we used the \textsc{vald} data base
\citep{1995A&AS..112..525P,1999A&AS..138..119K,1999PhST...83..162R} as
a source of atomic line parameters for opacity calculations and
abundance determination.\\ We first derived the effective temperature
\Teff\ by imposing simultaneously the Ti\,\textsc{i} excitation
equilibrium and the Ti ionization equilibrium and by fitting synthetic
fluxes to the available photometry (see Section \ref{subsec.sed}). We
adopted Ti because it is the only atom for which we measured a
significant number of lines, with reliable \loggf\ values, of two
ionization stages. In addition, the measured Ti\,\textsc{i} lines have
a much more uniform distribution in excitation energy, compared to any
other measured ion. We derived the line abundance from equivalent
widths analysed with a modified version \citep{1996ASPC..108..198T} of
the \textsc{width9} code \citep{1993KurCD..13.....K}.\\ We determined
\Teff\ assuming first a \logg\ value of 4.7 (typical of a
main-sequence mid-K-type star) and a microturbulence velocity (\vmic)
of 0.85\,km\,s$^{-1}$ \citep{2005ApJS..159..141V}. We could make this
first assumption because in this temperature regime both equilibria
(excitation and ionization) and the SED are almost completely
insensitive to \logg\ and \vmic\ variations
\citep{2012MNRAS.419L..34M}. We finally obtained and adopted an
effective temperature of 4400$\pm$200\,K.\\ Having fixed \Teff, and
knowing that \logg\ variations have negligible effects on
\vmic\ determination, we measured the microturbulence velocity by
imposing the equilibrium between the line abundance and equivalent
widths for Ti\textsc{i}, obtaining 0.8$\pm$0.3\,\kms, in agreement
with what is commonly measured for late-type stars
\citep{2012MNRAS.422..542P}.\\ In order to derive the metallicity of
POTS-1, we measured a total of 54 Fe\,\textsc{i} lines from which we
obtained an average abundance, relative to the Sun
\citep{2009ARA&A..47..481A} of $[$Fe/H$]$\,=\,$-$0.03$\pm$0.15, where
the uncertainty takes into account the uncertainties on the
atmospheric parameters (see \citet{2009A&A...503..945F} for more
details). We measured also the abundance of Mg, Al, Si, Ca, Sc, Ti, V,
Cr, Mn, Co, Ni, Y, Zr and Ba, all of them being consistent with the
solar values. The fit of synthetic spectra to the observed spectrum
did not require the addition of any rotational broadening; therefore,
the stellar projected rotational velocity (\vsini) is constrained by
the minimum measurable \vsini\ allowed by the resolution of the
spectrograph.  With the given spectral resolution of 40000, \vsini\ is
$\leq$\,5.3\,\kms, in agreement with the derived evolutionary status
of the star.\\ In a completely independent analysis of the
spectroscopic data, we fitted synthetic models computed by the
\textsc{wita6} program \citep{1997Ap&SS.253...43P} for a grid of
\textsc{sam12} model atmospheres \citep{2003ARep...47...59P} of
different \Teff, \logg, $[$Fe/H$]$ and find stellar parameters which
are in very good agreement with the ones derived with the
\textsc{marcs} models.\\ Figs \ref{isochrones} and \ref{mass} show the
Dartmouth isochrones in the range from 1 to 13\,Gyr for solar
metallicity \citep{2008ApJS..178...89D}. The \logg\ and mass evolution
of a mid-K dwarf over the age of the Universe is rather small. Given
the possible range for \Teff\ , we constrain the surface gravity to
\logg\ \,=\,4.63$\pm$0.05 and the mass to
M$_{\star}$\,=\,0.695$\pm$0.050\,M$_{\odot}$. We compared these
results to the values obtained using low-mass stellar evolution models
from \citet{1998A&A...337..403B} which agree very well
(\logg\ \,=\,4.66$\pm$0.04 and
M$_{\star}$\,=\,0.70$\pm$0.06\,M$_{\odot}$).\\ As a spectroscopic
confirmation of the derived \logg\ value, Fig. \ref{fig:sodium} shows
a region of the summed spectra centred on the Na\,\textsc{i}\,D lines
together with synthetic spectra of three different surface gravities
which underlines our adopted value of \mbox{\logg\ \,=\,4.63}.\\
\begin{table*} 
  \centering 
  \begin{tabular}{cccccr} \hline 
    \multicolumn{1}{|c}{BJD} & \multicolumn{1}{|c}{Mode} & \multicolumn{1}{|c}{Phase} & \multicolumn{1}{|c}{S/N} & \multicolumn{1}{|c}{RV} & \multicolumn{1}{|c}{Bisector span} \\ \hline
    2454924.70252   & Fibre   & 0.278      & 12.2    & --16.779$\pm$0.358\,km\,s$^{-1}$  &   0.112$\pm$0.382\,km\,s$^{-1}$  \\        
    2454924.73647   & Fibre   & 0.288      & 12.9    & --16.879$\pm$0.378\,km\,s$^{-1}$  & --0.001$\pm$0.229\,km\,s$^{-1}$  \\        
    2454924.77903   & Fibre   & 0.302      & 16.3    & --16.739$\pm$0.428\,km\,s$^{-1}$  &   0.024$\pm$0.197\,km\,s$^{-1}$  \\        
    2454924.81228   & Fibre   & 0.313      & 16.2    & --16.476$\pm$0.578\,km\,s$^{-1}$  & --0.128$\pm$0.228\,km\,s$^{-1}$  \\        
    2455066.49230   & Fibre   & 0.139      & 12.6    & --17.252$\pm$0.532\,km\,s$^{-1}$  & --0.060$\pm$0.227\,km\,s$^{-1}$  \\        
    2455258.62018   & Fibre   & 0.928      & 20.1    & --16.289$\pm$0.552\,km\,s$^{-1}$  &   0.304$\pm$0.209\,km\,s$^{-1}$  \\        
    2455289.79078   & Slit    & 0.790      & 14.8    & --15.976$\pm$0.401\,km\,s$^{-1}$  &   0.000$\pm$0.152\,km\,s$^{-1}$  \\        
    2455327.63875   & Slit    & 0.765      & 15.6    & --15.871$\pm$0.329\,km\,s$^{-1}$  &   0.117$\pm$0.143\,km\,s$^{-1}$  \\        
    2455349.61383   & Slit    & 0.718      & 12.5    & --15.629$\pm$0.569\,km\,s$^{-1}$  & --0.326$\pm$0.180\,km\,s$^{-1}$  \\        
    2455376.49998   & Slit    & 0.224      & 11.9    & --15.941$\pm$0.354\,km\,s$^{-1}$  & --0.156$\pm$0.163\,km\,s$^{-1}$  \\        
    2455384.53550   & Slit    & 0.767      & 16.4    & --15.512$\pm$0.514\,km\,s$^{-1}$  &   0.259$\pm$0.196\,km\,s$^{-1}$  \\        
    2455387.52652   & Slit    & 0.713      & 14.3    & --15.947$\pm$0.308\,km\,s$^{-1}$  & --0.031$\pm$0.192\,km\,s$^{-1}$  \\        
    2455387.56369   & Slit    & 0.725      & 13.2    & --15.701$\pm$0.304\,km\,s$^{-1}$  &   0.058$\pm$0.161\,km\,s$^{-1}$  \\ \hline 
  \end{tabular}
  \caption{Overview of the spectroscopic observations of POTS-1 taken
    with UVES. The first four columns give the Barycentric Julian Date
    of the exposure mid-point, the instrument mode, the planet's
    orbital phase at the time of observation and the average
    signal-to-noise ratio of the spectra per resolution element.
    Columns 5 and 6 give the RV and the bisector span measurements.}
  \label{tab.RV_epochs}
\end{table*} 
\begin{table*} 
  \centering 
  \begin{tabular}{rcl}\hline
    \Teff           &=& 4400$\pm$200\,K                   \\
    \logg           &=& 4.63$\pm$0.05                     \\
    $[$Fe/H$]$      &=& --0.03$\pm$0.15                   \\
    \vmic           &=& 0.8$\pm$0.2\,\kms                 \\
    \vsini      &$\le$& 5.3\,\kms                         \\
    $M_{s}$         &=& 0.695$\pm$0.050\,M$_{\rm{\odot}}$ \\
    $K$             &=& 407$\pm$117\,m\,s$^{-1}$          \\
    $M_{p}$         &=& 2.31$\pm$0.77\,M$_{\rm{Jup}}$     \\ \hline
  \end{tabular}
  \caption{Stellar parameters of POTS-1 and RV amplitude as determined
    from our spectroscopic observations and SED fitting.}
  \label{RV_parameters}
\end{table*} 
\subsubsection{Spectral energy distribution}
\label{subsec.sed}
\begin{figure}
  \begin{center}
    \includegraphics[width=0.45\textwidth]{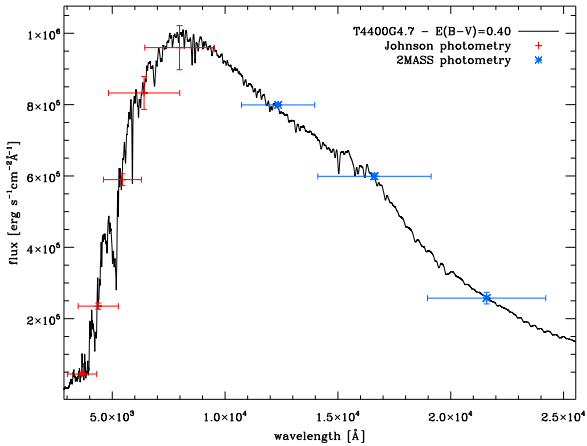}
    \caption{SED fitting for POTS-1. The plot shows the comparison
      between \textsc{marcs} model theoretical fluxes calculated with
      \Teff\,=\,4400\,K and log\,$g$\,=\,4.7 (our final set of
      fundamental parameters), taking into account a reddening of
      E($B$-$V$)\,=\,0.40\,mag, with the Johnson--Cousins $UBVRI$
      photometry (red crosses) and 2MASS photometry (blue asterisks).}
    \label{fig:sed}
  \end{center}
\end{figure}

\noindent We extracted the $UBVRI$ magnitudes of POTS-1 from the
photometric calibration (see Section
\ref{subsec.characterization}). In addition, we cross-matched with the
2MASS All-Sky Point Source Catalog (PSC) \citep{2006AJ....131.1163S}
and identified POTS-1 as object 13342613--6634520 and extracted the
$JHK$ magnitudes. All magnitudes are listed in Table
\ref{tab.POTS-C1}.\\ Using the optical and NIR photometry, we derived
\Teff\ making use of the \textsc{marcs} synthetic fluxes
\citep{2008A&A...486..951G}. Without accounting for any reddening, it
was not possible to fit POTS-1's SED assuming any \Teff, when using
simultaneously both the WFI and 2MASS photometry. By adding the
reddening to the SED, we derived a best-fitting temperature of
\Teff\,=\,4400$\pm$200\,K and reddening of
$E(B-V)$\,=\,0.40$\pm$0.05\,mag, where the optical photometry provided
the strongest constraint to the reddening while the NIR photometry
constrained mostly \Teff. Fig. \ref{fig:sed} shows our best fit. It is
important to mention that at these temperatures, the SED is only very
little dependent on the other atmospheric parameters of the star, such
as \logg\ or metallicity. Our best-fitting reddening value is in good
agreement with that obtained from galactic interstellar extinction
maps by \citet{2005AJ....130..659A}.\\ Given the reddening and the
galactic coordinates of the star, we derived a distance to the star of
1.2$\pm$0.6\,kpc. To independently check our results, we calculated
the distance to the star from the $V$ magnitude (corrected for the
reddening), the effective temperature (4400\,K) and a typical
main-sequence radius of 0.67\,R$_{\odot}$, obtaining a distance of
about 0.9\,kpc, in agreement with what was previously derived.\\
\begin{table*}
  \centering
  \begin{minipage}{\textwidth}
    \centering
    \begin{tabular}{|c|c|} \hline 
      \multicolumn{2}{|c|}{POTS-1 (2MASS 13342613--6634520)}                  \\ \hline
      RA   (J2000.0)        & 13$^{\rm{h}}$34$^{\rm{m}}$26$\fs$1              \\
      Dec. (J2000.0)        & --66$^\circ$34$'$52$''$                         \\
      $U$                   & 20.89$\pm$0.10\,mag                             \\
      $B$                   & 19.54$\pm$0.04\,mag                             \\
      $V$                   & 17.94$\pm$0.03\,mag                             \\
      $R$                   & 17.01$\pm$0.06\,mag                             \\
      $I$                   & 16.14$\pm$0.07\,mag                             \\
      $J$                   & 15.17$\pm$0.06\,mag                             \\
      $H$                   & 14.38$\pm$0.04\,mag                             \\
      $K$                   & 14.24$\pm$0.08\,mag                             \\
      $d$                   & 1.2$\pm$0.6\,kpc                                \\
      $E(B-V)$              & 0.40$\pm$0.05\,mag                              \\
      \Teff\                & 4400$\pm$200\,K                                 \\
      Spectral type         & K5V                                             \\ \hline 
    \end{tabular} 
  \end{minipage}
  \caption{Basic parameters of the POTS-1 system.}
  \label{tab.POTS-C1}
\end{table*}
\\
To confirm that POTS-1 hosts a planet, it is crucial to completely
exclude that it is a giant. The sole constraint given by the
Na\,\textsc{i}\,D line profile is not enough to guarantee that the
star is not a giant, as these lines could be fitted by various sets of
\Teff\ and \logg, i.e. lower gravity and lower temperature such as
\Teff\,=\,3950\,K and log\,$g$\,=\,3.0. However, to fit the SED with
such a set of stellar parameters, we would require a reddening much
smaller than 0.40\,mag. If we now assume that the star is a giant,
because of the large radius, the star would be at a distance much
larger than 1.2\,kpc,\footnote{A K giant has a minimum radius of
  $\sim$6\,R$_{\odot}$ which would require a distance of
  $\sim$7\,kpc.} implying a reddening much larger than 0.40\,mag, in
contradiction with what was required to fit the SED. Note that a large
radius of POTS-1 can also be excluded from the mean stellar density
obtained in the light-curve fitting (see below).\\ Table
\ref{RV_parameters} lists all host parameters that we derived from the
combined analysis of the spectroscopic data and SED fitting.\\
\subsubsection{RV analysis}
\label{subsec.spectroscopic3}
\begin{figure}
  \centering \includegraphics[width=0.45\textwidth]{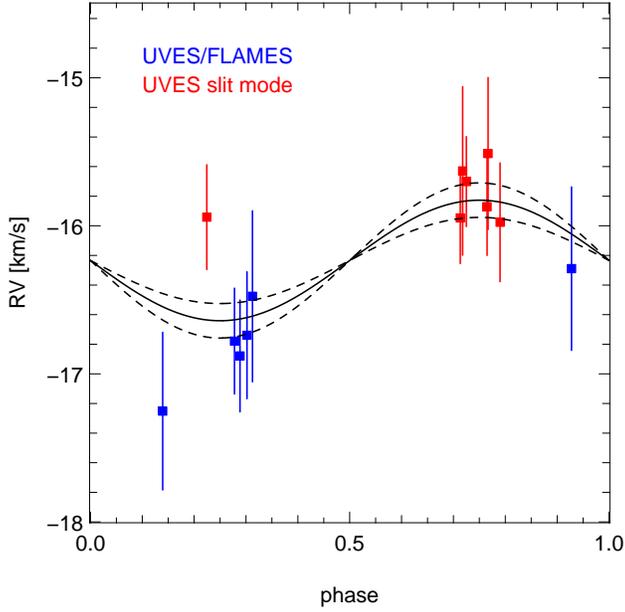} 
  \caption{The RV measurements of POTS-1 as a function of orbital
    phase. The solid and dashed lines show the best-fitting velocity
    amplitude and 1$\sigma$ upper and lower limits of the determined
    RV amplitude.}
  \label{rv}
\end{figure}

\begin{figure}
  \centering \includegraphics[width=0.45\textwidth]{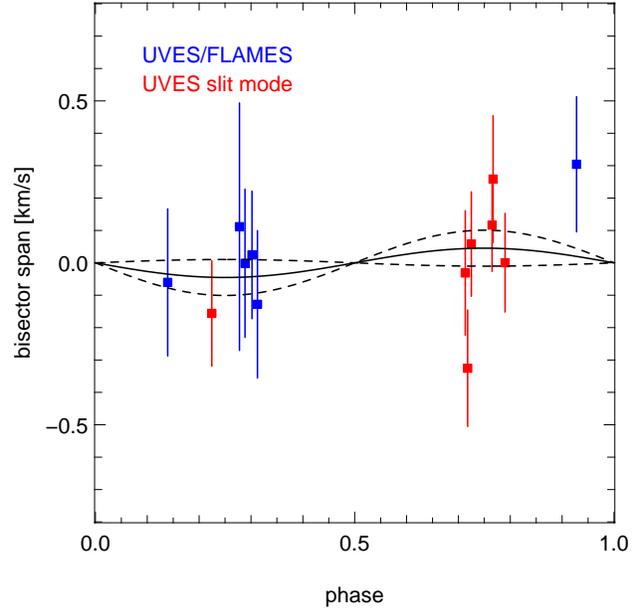}
  \caption{Bisector variations as a function of orbital phase. The
    solid line and dashed lines in the top panel indicate the
    least-squares fit to the sinusoidal variation in the bisector span
    and its uncertainty at 0.004$\pm$0.012 km\,s$^{-1}$.}
  \label{bisector}
\end{figure}

\begin{figure}
  \centering \includegraphics[width=0.45\textwidth]{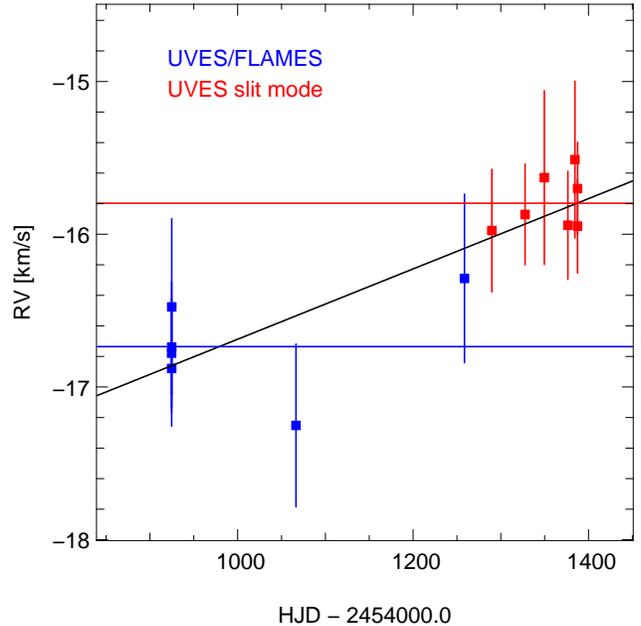}
  \caption{The RV measurements of POTS-1 as a function of BJD. A
    linear trend which could be caused by an undetected stellar
    companion fits the data quite well. In this case, the 1$\sigma$
    upper limit for the mass of the planet would be
    0.87\,M$_{\rm{Jup}}$ as measured by a least-squares fit to the
    residuals.}
  \label{trend}
\end{figure}

\noindent To calculate the RV variations of POTS-1 caused by the
transiting object, we cross-correlated the orders of the 13 spectra
with a Kurucz synthetic model spectrum for a star with
\Teff\,=\,4500\,K and
\logg\ \,=\,4.5\footnote{http://kurucz.harvard.edu/grids.html} (the
model grid has a resolution of 250\,K and 0.5\,dex). The resulting
barycentric corrected RV measurements are presented in Table
\ref{tab.RV_epochs}.  The uncertainties have been estimated from the
variation of the RV estimates obtained for the different echelle
orders. The final RV measurements as a function of orbital phase are
shown in Fig. \ref{rv}. We fitted the data with a sine function with
amplitude and zero-point velocity as free parameters. We find a
best-fitting RV amplitude of K\,=\,407$\pm$117\,m\,s$^{-1}$, with
V$_0$\,=\,--16.234\,km\,s$^{-1}$. Using the period obtained from the
transit fit, the RV curve fit and the stellar parameters derived
above, we obtain a planetary mass of
M$_{p}$\,=\,2.31$\pm$0.77\,M$_{{\rm{Jup}}}$.\\ In order to check for
line asymmetries, we determined the values of the bisector span
\citep[following][]{2001A&A...379..279Q} as a function of orbital
phase which are shown in Fig. \ref{bisector}. A least-squares fit of
the bisector span measurements with a sine function revealed no
significant variations at a level of
0.045$\pm$0.056\,km\,s$^{-1}$. Although this means that there is no
indication that the measured RV variations are due to line-shape
variations, caused by either stellar activity or blends, the errors
are very large, making any claim based on the bisector span
uncertain. \\ Due to the specific scheduling of the observations, most
of the measurements around phase $\sim$0.25 are taken in 2009 and most
of the observations around phase $\sim$0.75 in 2010. As a consequence,
the variation of the RV of POTS-1 can also be explained by a linear
trend which could be caused by a long-period massive companion to
POTS-1. Fig. \ref{trend} shows the distribution of the RV measurements
in time together with the best-fitting linear trend.  Although there
is no indication that a long-period companion to POTS-1 exists, we
explore this scenario by removing the linear trend and fitting a
sinusoidal RV variation to the residuals. The resulting semi-amplitude
is K\,=\,35$^{+151}_{-35}$\,m\,s$^{-1}$ which corresponds to a
planetary mass of
\mbox{M$_{p}$\,=\,0.20$^{+0.67}_{-0.20}$\,M$_{\rm{Jup}}$}.\\ Alternatively,
the detected RV variation can be explained by an offset between the
UVES slit-mode and fibre-mode measurements. RV offsets of up to a few
100 m\,s$^{-1}$ are commonly found when combining RV observations from
different instruments \citep[see e.g.][]{2012A&A...545A.139P}. The
solid red and blue lines in Fig. \ref{trend} show the average of all
measurements taken with UVES in slit mode and fibre mode,
respectively. Fitting a sinusoid to the RV measurements of the
observations taken in different modes independently we find a
semi-amplitude of K\,=\,337$^{+203}_{-202}$\,m\,s$^{-1}$ for the
fibre-mode obesrvations and a semi-amplitude of
K\,=\,67$^{+142}_{-67}$\,m\,s$^{-1}$ for the slit-mode observations
which translates into mass estimates of
M$_{p}$\,=\,1.91$^{+1.30}_{-1.19}$ and
0.38$^{+0.86}_{-0.38}$\,M$_{\rm{Jup}}$, respectively.\\
\subsection{Photometric transit observations with GROND} 
\label{subsec.photometric} 
\begin{figure}
  \centering \includegraphics[width=0.32\textwidth]{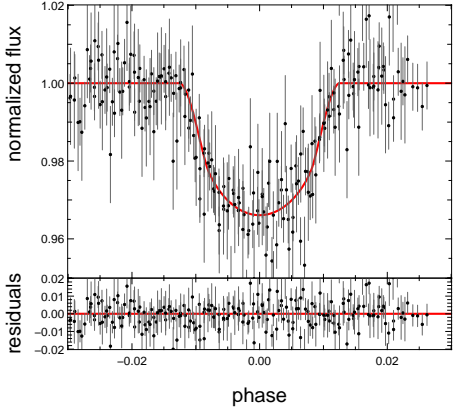} 
  \caption{Folded $g'$-band light curve of POTS-1 containing the
    observations of three individual transits. The red line shows the
    best-fitting model with the parameters listed in Table
    \ref{parameters}.}
  \label{trans_first}
\end{figure}

\begin{figure}
  \centering \includegraphics[width=0.4\textwidth]{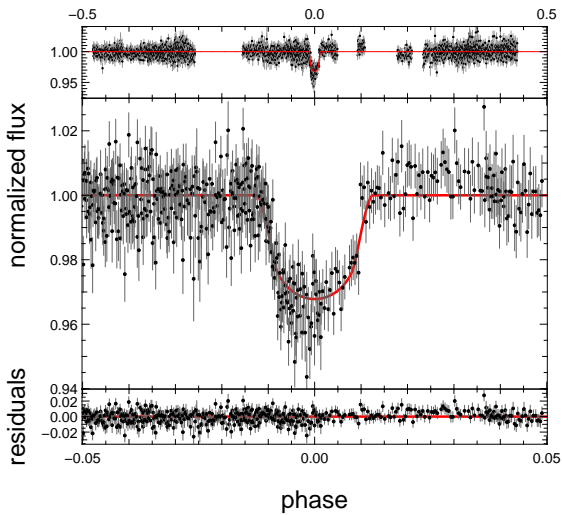} 
  \caption{Folded POTS R-band light curve of POTS-1. The red line
    shows the best-fitting model with the parameters listed in Table
    \ref{parameters}. Note that due to a deficiency of points at phase
    0.5, the R-band light curve would be consistent with a period that
    is actually half the detected period (i.e. with each second
    eclipse being hidden in the gap). However, we ruled this out with
    additional GROND observations at phase 0.5 which showed no sign of
    the eclipse.}
  \label{fig.POTS-C1}
\end{figure}

In order to precisely determine the physical parameters of the POTS-1
system such as period, $t_0$, planet radius and orbital inclination,
we observed three full transits of POTS-1 with the GROND instrument
\citep{2008PASP..120..405G} mounted on the MPI/ESO 2.2m telescope at
the La Silla observatory. The GROND instrument has been used several
times for the confirmation of a transiting planet candidate
\citep[e.g.][]{2009A&A...497..545S} and for detailed follow-up studies
of known transiting extra-solar planets \citep[see
  e.g.][]{2010A&A...522A..29L,2012A&A...538A..46D,2012A&A...539A.159N}.
GROND is a seven-channel imager that allows us to take four optical
($g'$, $r'$, $i'$ and $z'$) and three near-infrared ($JHK$) exposures
simultaneously. For our observations, the $JHK$ light curves turned
out to have a large scatter; we therefore did not use them in our
analysis.\\ We observed POTS-1 during the nights of 2009 April 17,
2010 April 6 and 2010 July 13 and collected a total of 84, 68 and 123
images in each optical band. The exposure time was ranging from 133 to
160\,s resulting in a cycle rate of about 3.5\,min.\\ All optical
images were reduced with the \textsc{mupipe} software developed at the
University Observatory in
Munich.\footnote{http://www.usm.lmu.de/$\sim$arri/mupipe} After the
initial bias and flat-field corrections, cosmic rays and bad pixels
were masked and the images were resampled to a common grid. The frames
did not suffer from detectable fringing, even in the z' band. We
performed aperture photometry on POTS-1 and typically 10 interactively
selected reference stars after which light curves were created for
each of the four bands.\\ In order to account for the wide range of
seeing conditions, the aperture radii were chosen between 4 and
16\,pixels (corresponding to 0.6--2.4\,arcsec) by minimizing the rms
scatter in the out of transit part of the light curves. For each band
at a given night, we used a fixed aperture size. The sky background
was determined as the median value in an annulus with an inner radius
of 25 and an outer radius of 30\,pixels measured from the object
centre positions.\\ The light curves were fitted with analytic models
presented by \citet{2002ApJ...580L.171M}. We used quadratic
limb-darkening coefficients taken from \citet{2011A&A...529A..75C},
for a star with metallicity $[\rm{Fe/H}]$\,=\,0.0, surface gravity
\mbox{log\,$g$\,=\,4.7} and effective temperature
\Teff\,=\,4400\,K. The values of the limb-darkening coefficients are
obtained as linear interpolations of the available grid.\\ Using a
simultaneous fit to the individual light curves in all five bands (the
four optical GROND bands plus the WFI R band) we derived the period
$P$, epoch $t_0$, mean stellar density
$\rho_s$\,=\,$M_{s}$\,/\,$R^3_{s}$ in solar units, the radius ratio
$R_{p}$\,/\,$R_{s}$ and the impact parameter $\beta_{\rm{impact}}$ (in
units of $R_{s}$). Together with three scaling factors for each GROND
filter (one for each night) and one scaling factor for the R-band
light curve, a total of 18 free parameters were fitted. The light
curves and the best-fitting models are shown in
Figs. \ref{trans_first}--\ref{fig.POTS-C1}. The rms values of the
residuals in the combined light curves are 8.9, 3.6, 3.8, 5.4 and
7.7\,mmag for $g'$, $r'$, $i'$, $z'$ and R bands,
respectively.\\ Using the stellar mass estimate of
$M_s$\,=\,0.695$\pm$0.050\,M$_{\odot}$ as derived in Section
\ref{subsec.spectroscopic}, we calculate the radius of the star and the
planet, as well as the inclination of the orbit. In order to derive
uncertainties of the system parameters, we minimized the $\chi^2$ on a
grid centred on the previously found best-fitting parameters and
searched for extreme grid points with $\Delta\chi^2$\,=\,1 when
varying one parameter while simultaneously minimizing over the
others. The resulting final parameters and error estimates are listed
in Table \ref{parameters}. Note that the best-fitting value of
\logg\ based on the light curves is in reasonable agreement with the
value derived from the spectra and colours.\\ We fit the individual
transits in order to search the light curves for transit timing
variations. The WFI R-band light curve shows only a single complete
transit. Together with the three transits observed with GROND, we find
the central transit times reported in Table \ref{TTV}. We find no
significant transit timing variations.\\
\begin{figure}
  \centering \includegraphics[width=0.32\textwidth]{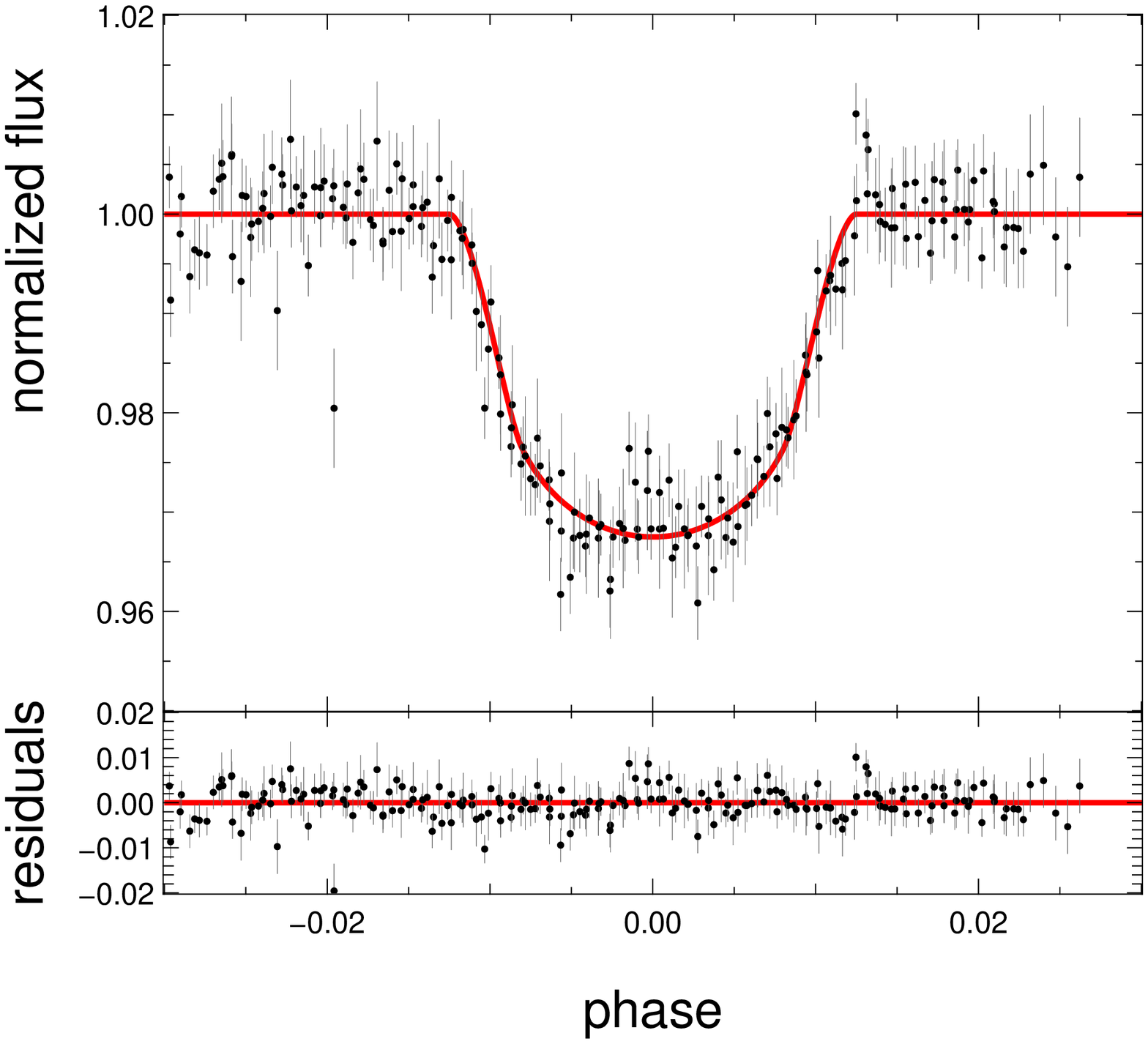} 
  \caption{Folded $r$'-band light curve of POTS-1 containing the
    observations of three individual transits. The red line shows the
    best-fitting model with the parameters listed in Table
    \ref{parameters}.}
\end{figure}
\begin{figure}
  \centering \includegraphics[width=0.32\textwidth]{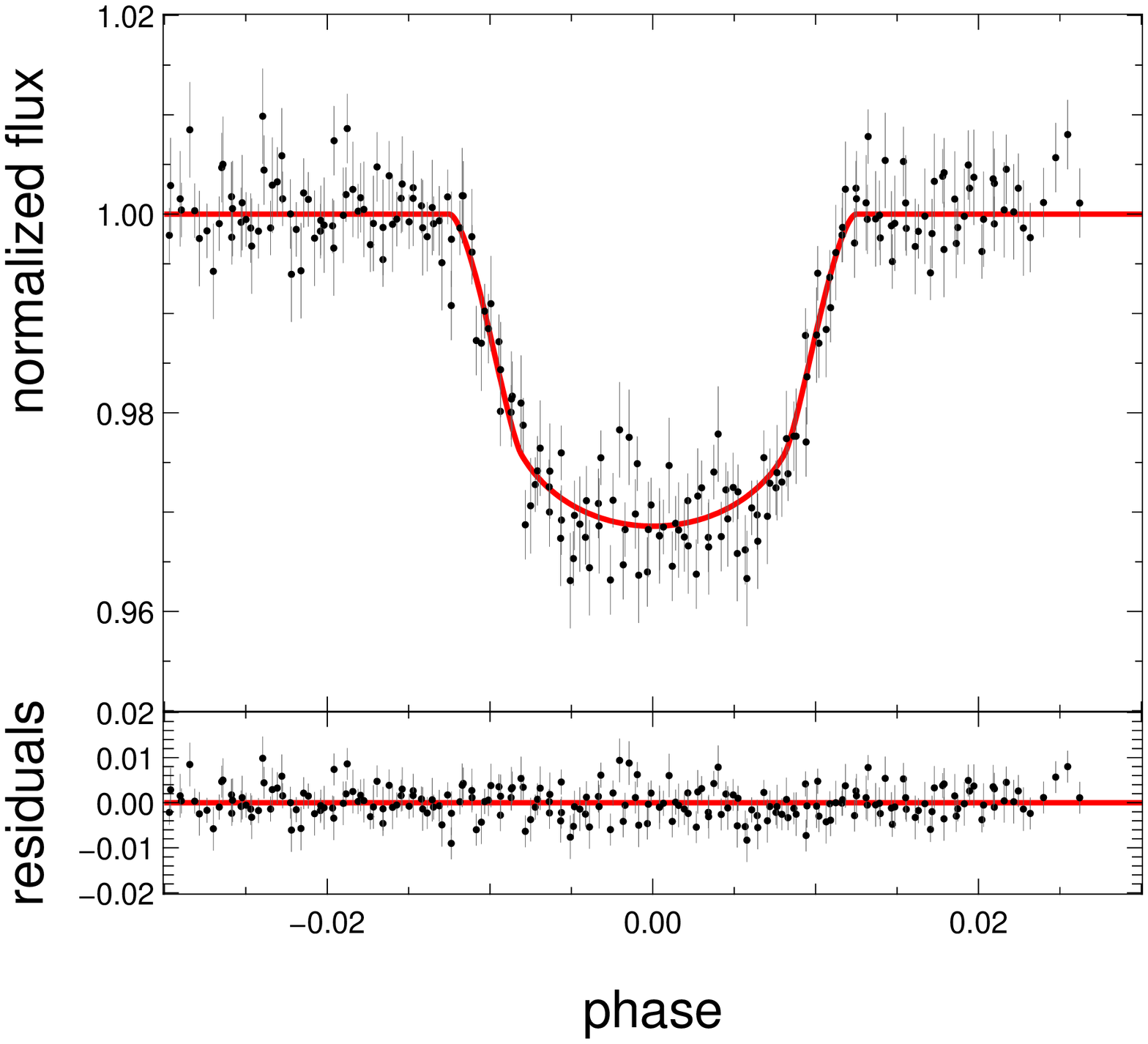} 
  \caption{Folded $i'$-band light curve of POTS-1 containing the
    observations of three individual transits. The red line shows the
    best-fitting model with the parameters listed in Table
    \ref{parameters}.}
\end{figure}
\begin{figure}
  \centering \includegraphics[width=0.32\textwidth]{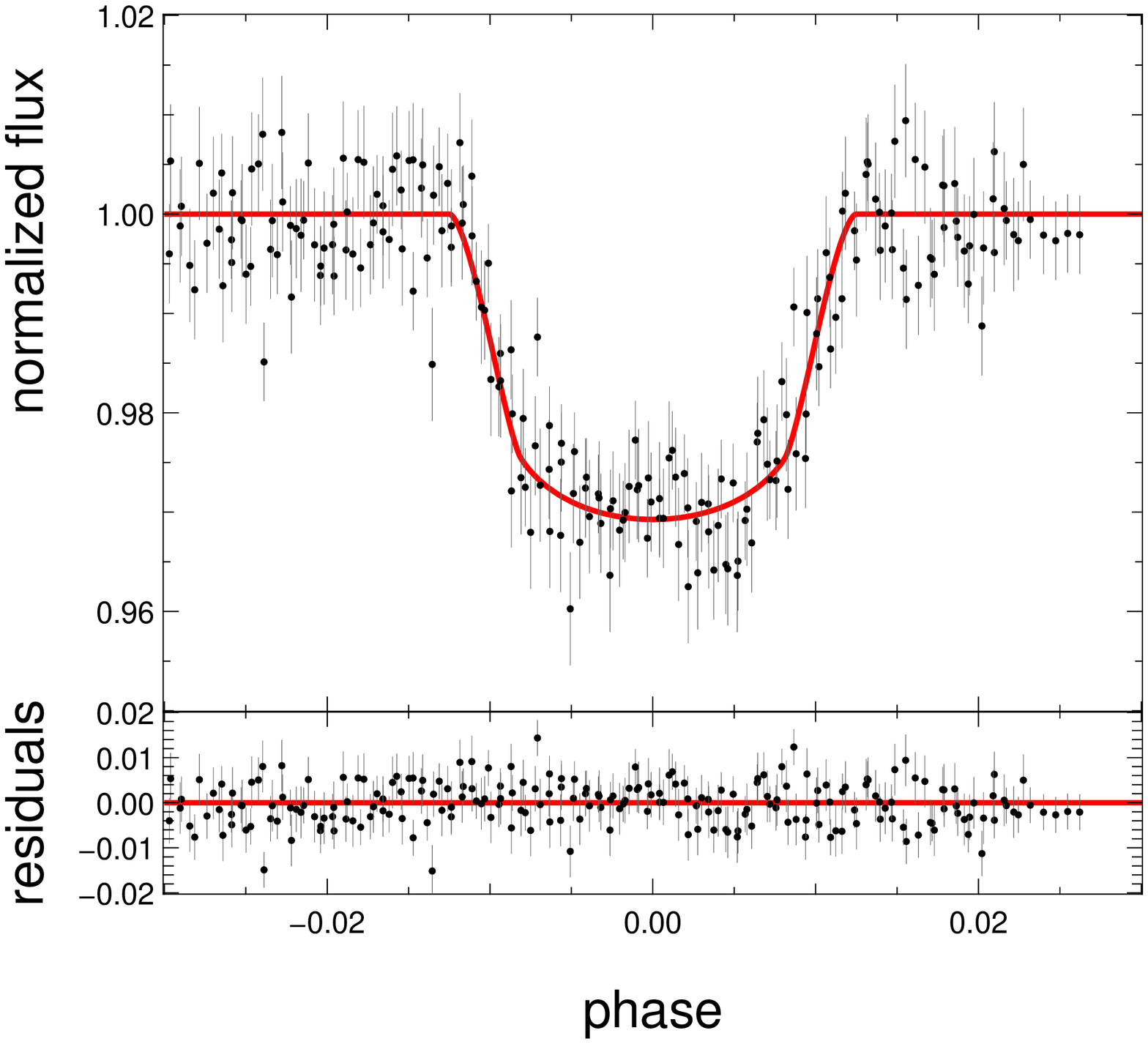} 
  \caption{Folded $z'$-band light curve of POTS-1 containing the
    observations of three individual transits. The red line shows the
    best-fitting model with the parameters listed in Table
    \ref{parameters}.}
\end{figure}
\begin{table*} 
  \centering 
  \begin{tabular}{rcl}\hline
    \vspace{1.3ex}$\rho_s$              &=& 3.417$^{+0.410}_{-0.261}$\,$\rho_{\rm{\odot}}$ \\
    \vspace{1.3ex}$R_{p}$\,/\,$R_{s}$   &=& 0.16432$^{+0.00176}_{-0.00218}$                \\
    \vspace{1.3ex}$\beta_{\rm{impact}}$ &=& 0.459$^{+0.045}_{-0.073}$                      \\
    \vspace{1.3ex}$t_0$                 &=& 2454231.65488$\pm$4.4$\times10^{-4}$\,BJD      \\
    \vspace{1.3ex}$P$                   &=& 3.16062960$\pm$1.57$\times10^{-6}$\,d          \\
    \vspace{1.3ex}$i$                   &=& 88$\fdg$06$^{+0.47}_{-0.34}$                   \\
    \vspace{1.3ex}$R_{p}$               &=& 0.941$^{+0.036}_{-0.047}$\,R$_{\rm{Jup}}$      \\
    \vspace{1.3ex}$R_{s}$               &=& 0.588$^{+0.020}_{-0.028}$\,R$_{\odot}$         \\
    \vspace{1.3ex}\logg\                &=& 4.74$\pm$0.07                                  \\
    \vspace{1.3ex}$a_{p}$               &=& 0.03734$\pm$0.00090\,AU                        \\\hline
  \end{tabular}
  \caption{The orbital, host star and planetary parameters of the
    POTS-1 system as determined from the photometric observations.}
  \label{parameters}
\end{table*} 
\begin{table*} 
  \centering 
  \begin{tabular}{cccr}\hline 
    \multicolumn{1}{|c}{Civil date} & \multicolumn{1}{|c}{Filter} & \multicolumn{1}{|c}{$t_0$ (BJD)} & \multicolumn{1}{|c}{O--C} \\\hline 
    14.03.2007 & $R$      & 2454174.76322$\pm$7.0$\times10^{-4}$ & --3.3$\times10^{-4}$ \\
    17.04.2009 & $griz$   & 2454939.63572$\pm$2.9$\times10^{-4}$ & --1.9$\times10^{-4}$ \\
    06.04.2010 & $griz$   & 2455293.62673$\pm$3.8$\times10^{-4}$ &   3.0$\times10^{-4}$ \\
    13.07.2010 & $griz$   & 2455391.60583$\pm$3.7$\times10^{-4}$ & --1.1$\times10^{-4}$ \\\hline 
  \end{tabular} 
  \caption{Central times $t_0$ for the individual transits observed
    with WFI and GROND. No significant transit timing variations are
    detected.}
  \label{TTV}
\end{table*} 
\subsection{High-resolution imaging with NACO}
\label{subsec.imaging}
\begin{figure}
  \centering
  \includegraphics[width=0.45\textwidth]{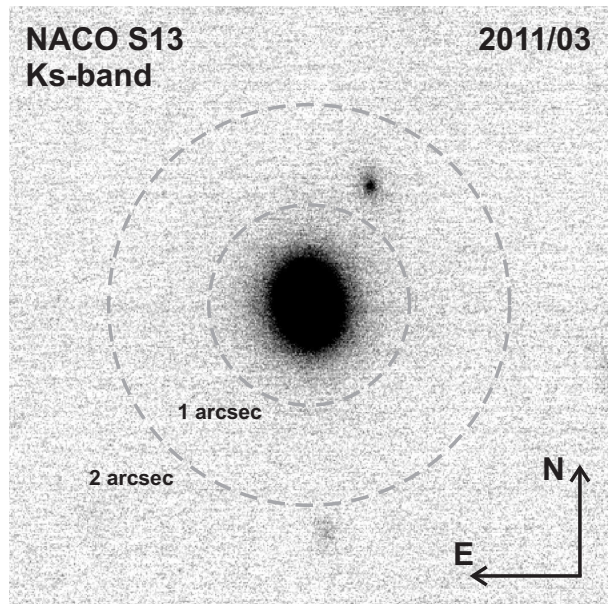}
  \caption{Fully reduced NACO image of POTS-1. The closest source
    found next to POTS-1 located about 1.3\,arcsec north-west.}
  \label{NACO1}
\end{figure}

\begin{figure}
  \centering
  \includegraphics[width=0.45\textwidth]{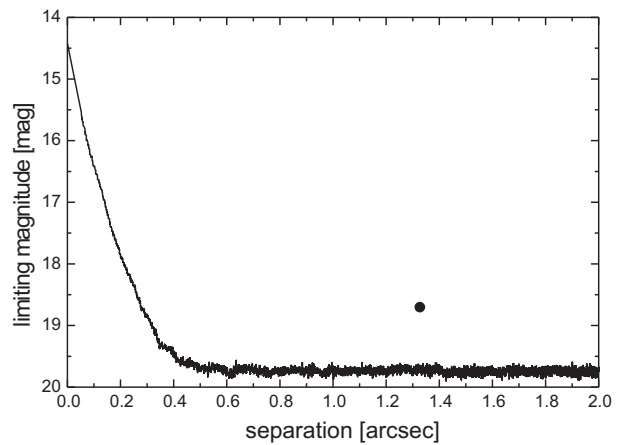}
  \caption{The achieved detection limit of our NACO observation is
    about 20\,mag at an angular separation of 0.5\,arcsec from
    POTS-1. Within 1\,arcsec around the star, there is no object
    detected. The filled circle indicates the position and brightness
    of the closest source found next to POTS-1 (compare with
    Fig. \ref{NACO1}).}
  \label{NACO2} 
\end{figure}

\begin{figure}
  \centering
  \includegraphics[width=0.45\textwidth]{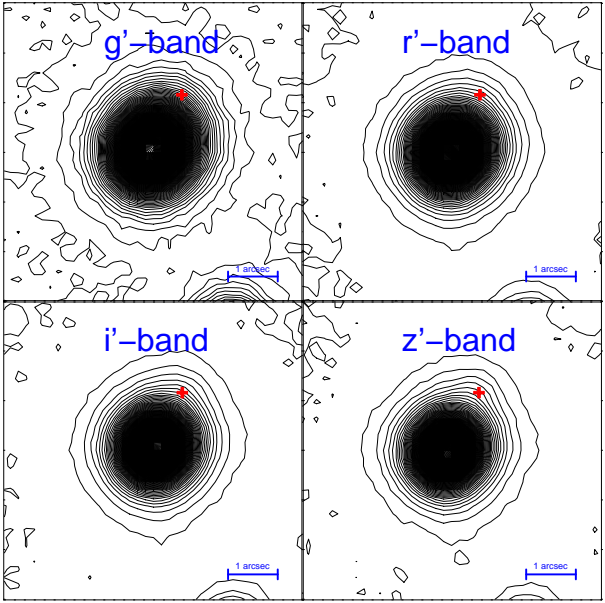}
  \caption{Isophotal contours for a stack of the 20 best seeing images
    in each of the four optical GROND bands. The 150 contours are
    equally spaced between 0 and the peak values of POTS-1. The object
    that is located 1.3\,arcsec north-west (indicated by the red
    cross) is slightly distorting the contours of the redder bands
    $i'$ and $z'$ but not visible in the $g'$ and $r'$ bands.}
  \label{contours} 
\end{figure}

On 2011 March 25 we obtained follow-up high-contrast imaging
observations of POTS-1 with NACO
\citep{2003SPIE.4841..944L,2003SPIE.4839..140R}, the AO imager of
ESO-VLT, at the Paranal observatory in Chile. The wavefront analysis
was performed with NACO's visible wavefront sensor VIS, using the AO
reference star 2MASS J13342468--6634348 (V$\sim$15.1\,mag,
$K_s$$\sim$9.3\,mag), which is located about 19.2\,arcsec north-west
of POTS-1.\\ The observations were carried out at an airmass of about
1.35 with a seeing of $\sim$0.8\,arcsec and coherence time
$\tau_{0}\sim$3\,mas of the atmosphere in average, which corresponds
to a value of the coherent energy of the PSF of the star of 33$\pm$4
per cent.\footnote{As determined by NACO's RTC.}\\ We took 21 frames
with NACO's high-resolution objective S13 (pixel scale
13.23\,mas\,pixel$^{-1}$ and 13.5\,arcsec\,$\times$\,13.5\,arcsec FoV)
in the $K_s$ band each with a detector integration time of 60\,s, in
the HighDynamic detector mode. For the subtraction of the bright
background of the night sky in the $K_s$ band, the telescope was moved
between individual integrations (standard jitter mode) with a jitter
width of 7\,arcsec, sufficiently large to avoid overlapping of the PSF
of detected sources.\\ For the flat-field correction, internal lamp
flats as well as skyflats were taken before and after the observations
during daytime or twilight, respectively. The data reduction
(background estimation, background subtraction, and then
flat-fielding), as well as the final combination (shift+add) of all
images, was then performed with \textsc{eso-eclipse}\footnote{ESO C
  Library for an Image Processing Software Environment.}
\citep{2001ASPC..238..525D}.\\ Our fully reduced NACO image of POTS-1
is shown in \mbox{Fig. \ref{NACO1}}. The elliptical shape of the PSF
is due to the fact that we observed the object outside the isoplanatic
angle. This was necessary because POTS-1 is too faint to be used as AO
reference and we had to choose a brighter nearby reference. A careful
look at Fig. \ref{NACO1} shows that the long axis of the elliptical
PSF (PA$\sim$5$^\circ$) is not aligned with the direction to the AO
reference star (PA$\sim$330$^\circ$) which could be an indication that
the ellipticity is caused by a second object very close to
POTS-1. However, the analysis of the PSF of the two brightest other
stars in the FoV revealed almost identical position angles,
elongations and ellipticities. We conclude that the ellipticity of the
NACO PSF of POTS-1 is not due to a close-by contaminating star but a
result of limited AO performance.\\ The achieved detection limit of
our NACO observation is illustrated in Fig. \ref{NACO2}. We reach a
detection limit of about 20\,mag at an angular separation of
0.5\,arcsec from POTS-1. Within 1\,arcsec around the star, there is no
object detected. The closest source found next to POTS-1 is located
about 1.3\,arcsec north-west from the star and exhibits a magnitude
difference of $\Delta K_s$\,=\,4.28$\pm$0.08\,mag. If physically
bound, the system would be similar to \citet{2006ApJ...641L..57B},
however with a much larger separation of $\sim$1500\,AU.\\ Using the
Besan\c{c}on model \citep{2003A&A...409..523R} of the POTS target
field, we estimate the by-chance alignment of a K\,=\,14.2\,mag star
and a star with $\Delta$K\,$\le$\,4.3\,mag and a separation
$\le$\,1.3\,arcsec to be $\sim$24 per cent.\\ In order to check if
this star could result in a significant blending of the light curves
that were obtained with GROND (see Section \ref{subsec.photometric}),
we built a stack of the 20 best seeing images in each of the four
bands $g'$, $r'$, $i'$ and $z$'. Fig. \ref{contours} shows isophotal
contours of these images together with the position of the
contaminating source. Since the two bluer bands $g'$ and $r'$ are less
affected (if at all) than the $i'$ and $z'$ bands we conclude that the
contaminating object is redder than POTS-1 and the contamination of
the light curves due to this object is less than 2 per cent in the
optical bands and therefore has a negligible impact on the parameters
we derived for the POTS-1 system.\\
\subsection{Rejection of blend scenarios}
\label{L9.blend}
\begin{figure}
  \centering \includegraphics[width=0.45\textwidth]{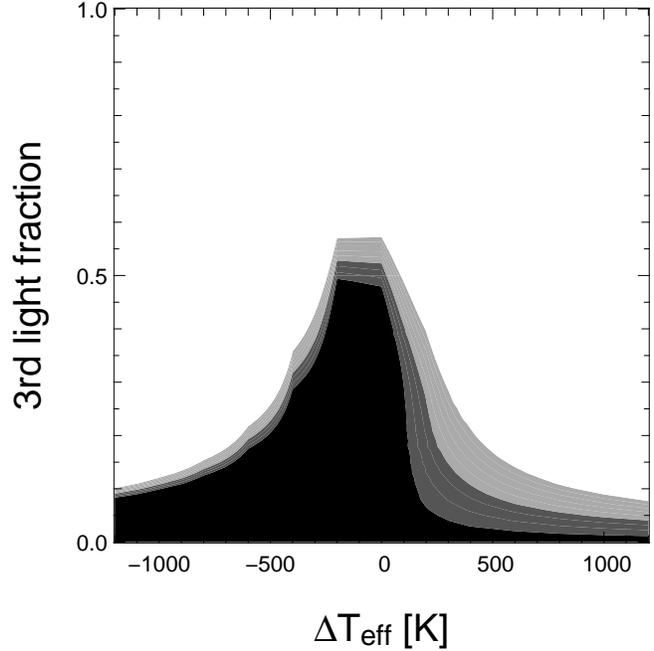}
  \caption{1$\sigma$, 2$\sigma$ and 3$\sigma$ contours for all
    possible blend scenarios with a light contribution coming from the
    contaminating object quantified as the average third light
    fraction in all five available bands. Values of f$_{3rd}>$\,50 per
    cent correspond to scenarios where the blending object is brighter
    than the eclipsing system, and values of f$_{3rd}<$\,50 per cent
    correspond to scenarios where the eclipsing system is
    dominant. For effective temperature differences between POTS-1 and
    the blending star that are larger than a few hundred K, we can
    rule out a significant contamination. For very similar spectral
    types, up to $\sim$50 per cent contamination is still consistent
    with the light curves.}
  \label{blending}
\end{figure}

Although the small amplitude of the RV variation, the bisector
analysis presented in Section \ref{subsec.spectroscopic} and the
high-resolution imaging presented in Section \ref{subsec.imaging}
confirm the planetary nature of POTS-1b, we perform an additional test
to strengthen our result. Based on the light curves we rule out a
scenario in which the POTS-1 system consists of an eclipsing binary
system which is blended by a third object that is coincidentally in
the line of sight or physically bound to the binary. In such a blend
system, the transit depth would vary across the different bands if the
spectral type of the blending star differs from the spectral type of
the binary, since in each band the blending fraction will be
different. If, on the other hand, the blending star is of a very
similar spectral type, the shape of the transit would become
inconsistent with the observations.\\ In a quantitative analysis
similar to \citet{2011ApJ...727...24T}, we simulated light curves of a
large range of possible blend scenarios, i.e. foreground/background
eclipsing binary and physical triplet systems with different amount of
blending -- quantified as the average fraction of third light in the
observed bands -- and different spectral types of the blending
sources. To each of the blended light curves, we fitted an analytic
eclipse model according to the equations of
\citet{2002ApJ...580L.171M} with quadratic limb-darkening coefficients
taken as linear interpolations of the grid published by
\citet{2011A&A...529A..75C} and compared the $\chi^2$ of the blended
to the unblended case. Fig. \ref{blending} shows the 1$\sigma$,
2$\sigma$ and 3$\sigma$ contours for all possible scenarios. For a
small difference in $T_{eff}$ between POTS-1 and the blending source,
an average third light fraction of up to $\sim$50 per cent is
consistent with the observed light curves. Note however that in this
case the radius of POTS-1b would be $\sim$1.33\,R$_{\rm{Jup}}$ which
is still well comparable to the radii of other extra-solar planets
found in the past. In this sense, we can rule out all eclipsing binary
scenarios with a fore- or background star that is contributing more
than 50 per cent of the total light, which underlines the confirmation
of the planetary nature of POTS-1b. Note that although this analysis
shows that a blending light of up to 50 per cent is consistent with
the light curves, there is no indication at all for a blend. We are
therefore confident that all parameters derived in the previous
sections are correct.\\ We investigated the possibility to apply the
centroid-shift method \citep[see e.g.][]{2010ApJ...724.1108J} to the
GROND data as an additional test for a blend scenario; however, the
faintness of POTS-1 and the low number of available out-of-transit
points did not allow us to get any useful measurement.\\
\section{POTS-C2}
\label{sec.POTS-C2}
\begin{figure}
  \begin{center}
    \includegraphics[width=0.45\textwidth]{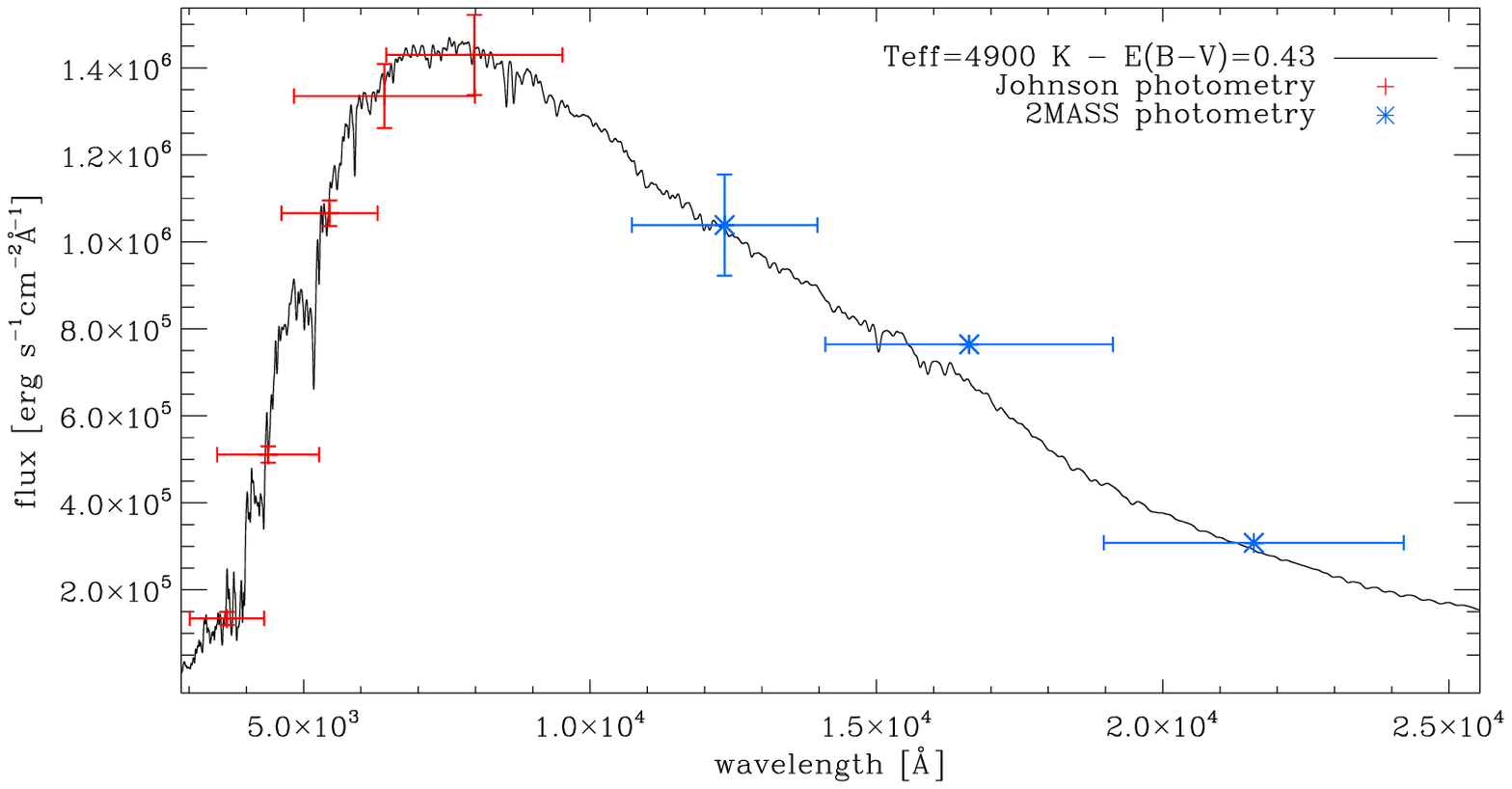}
    \caption{SED fitting for POTS-C2. The plot shows the comparison
      between \textsc{marcs} model theoretical fluxes calculated with
      \Teff\,=\,4900\,K and \logg\ =\,4.5 (our final set of
      fundamental parameters), taking into account a reddening of
      E($B$-$V$)\,=\,0.43\,mag, with the Johnson--Cousins $UBVRI$
      photometry (red crosses) and 2MASS photometry (blue asterisks).}
    \label{fig:sed2}
  \end{center}
\end{figure}

\begin{figure}
  \centering
  \includegraphics[width=0.4\textwidth]{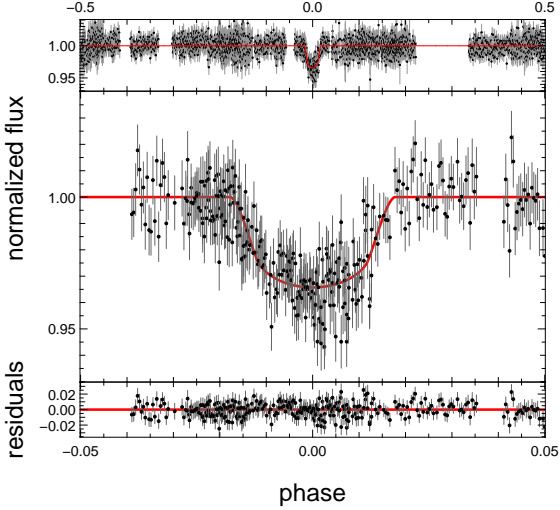}
  \caption{Folded R-band light curve of POTS-C2. The red line shows
    the best-fitting model with the parameters listed in \mbox{Table
      \ref{tab.POTS-C2}}.}
  \label{fig.POTS-C2}
\end{figure}

In this section, we present our second best candidate POTS-C2. We list
all $UBVRI$ magnitudes from the photometric calibration (see Section
\ref{subsec.characterization}) and the 2MASS $JHK$ magnitude
\citep{2006AJ....131.1163S} in Table \ref{tab.POTS-C2}. Since no H-
and K-band error estimates were available in the 2MASS PSC, we list
the typical 2MASS uncertainty for the given apparent
brightness.\\ Using the broad-band photometry, we derived the stellar
surface temperature \Teff\ making use of the synthetic fluxes
calculated with \textsc{marcs} models \citep{2008A&A...486..951G}.  We
were able to derive an effective temperature of
\Teff\,=\,4900$\pm$400\,K, a reddening of
\mbox{$E(B-V)$\,=\,0.43$\pm$0.08\,mag} and a distance of
2.3$\pm$1.0\,kpc (see Fig. \ref{fig:sed2}).\\ To determine the orbital
and planetary parameters, we fitted analytical transit light-curve
models according to the equations of \citet{2002ApJ...580L.171M} with
quadratic limb-darkening coefficients taken as linear interpolations
of the grid published by \citet{2011A&A...529A..75C}. We make use of
the spectral type estimate from SED fitting. The free parameters of
the fits were period $P$, epoch $t_0$, inclination $i$ and planetary
radius $R_{p}$. We also fitted a scale in order to renormalize the
out-of-transit part of the light curve \mbox{to 1}.\\ In order to
derive uncertainties of the system parameters, we minimized the
$\chi^2$ on a grid centred on the previously found best-fitting
parameters and searched for extreme grid points with
$\Delta\chi^2$\,=\,1 when varying one parameter and simultaneously
minimizing over the others. The best-fitting parameters and their
error estimates are listed in Table \ref{tab.POTS-C2}. We show the
folded light curves of POTS-C2 in Fig. \ref{fig.POTS-C2} together with
the best analytical model fit. Note that there are signs of a shallow
secondary eclipse at phase value 0.5 which could indicate a
significant brightness of POTS-C2b. However, the available data are
not precise enough to make any claim in this direction.\\
\begin{table*}
  \centering
  \begin{minipage}{\textwidth}
    \centering
    \begin{tabular}{|c|c|} \hline 
      \multicolumn{2}{|c|}{POTS-C2 (2MASS 13373207--6650403)}                                   \\ \hline
      \multicolumn{2}{|l|}{Stellar parameters:}                                                 \\ \hline
      RA   (J2000.0)         & 13$^{\rm{h}}$37$^{\rm{m}}$32$\fs$1                               \\
      Dec. (J2000.0)         & --66$^\circ$50$'$41$''$                                          \\
      $U$                    & 20.90$\pm$0.12\,mag                                              \\
      $B$                    & 19.90$\pm$0.04\,mag                                              \\
      $V$                    & 18.50$\pm$0.03\,mag                                              \\
      $R$                    & 17.70$\pm$0.06\,mag                                              \\
      $I$                    & 16.91$\pm$0.07\,mag                                              \\
      $J$                    & 15.93$\pm$0.12\,mag                                              \\
      $H$                    & 15.16$\pm$0.10\,mag\footnote{No error estimate available, we therefore list a typical 2MASS uncertainty for this magnitude.}\\
      $K$                    & 15.09$\pm$0.14\,mag$^a$                                          \\
      $d$                    & 2.3$\pm$1.0\,kpc                                                 \\
      $E(B-V)$               & 0.43$\pm$0.08\,mag                                               \\
      \Teff\                 & 4900$\pm$400\,K                                                  \\
      Spectral type          & K2V                                                              \\ \hline 
      \multicolumn{2}{|l|}{Planetary and orbital parameters:}                                   \\ \hline
      $P$                    & 2.76303$\pm$2.0$\cdot10^{-4}$\,d                                 \\
      $t_0$                  & 2454241.3652$\pm$0.0025\,BJD                                     \\
      $R_{p}$                & 1.30$\pm$0.21\,R$_{\rm{Jup}}$                                    \\
      $i$                    & 87$\fdg$3$\pm$2$\fdg$0                                           \\
      $a_{p}$                & 0.0349\,AU\footnote{Assuming $M_{\star}$\,=\,0.74\,$M_{\odot}$.} \\ \hline
      \multicolumn{2}{|l|}{Light-curve and detection parameters:}                               \\ \hline
      S/N                    & 38.0                                                             \\
      $q$                    & 0.031                                                            \\
      $\Delta F$/$F$         & 0.024                                                            \\
      No. of transits        & 3                                                                \\
      No. of transit points  & 128                                                              \\
      Baseline rms           & 0.0129                                                           \\ \hline
    \end{tabular} 
  \end{minipage}
  \caption{Planetary, orbital and stellar parameters of the POTS-C2
    system. Note that zero eccentricity has been assumed.}
  \label{tab.POTS-C2}
\end{table*}
\section{Conclusions}
\label{sec.conclusions}
We have presented the results of the POTS, a pilot project aiming at
the detection of transiting planets in the Galactic disc. Using the
difference imaging approach, we produced high-precision light curves
of 16000 sources with several thousands of data points and a high
cadence of $\sim$2\,min. All light curves are publicly
available.\footnote{www.usm.uni-muenchen.de/$\sim$koppenh/pre-OmegaTranS/}\\ A
detailed analysis of the POTS light curves revealed two very
interesting planet candidates. With extensive spectroscopic and
photometric follow-up observations, we were able to derive the
parameters of the POTS-1 system and confirm its companion as a planet
with a mass of 2.31$\pm$0.77\,M$_{\rm{Jup}}$ and a radius of
0.94$\pm$0.04\,R$_{\rm{Jup}}$. This would lead to the conclusion that
POTS-1b has a relatively high but not unusual density.\footnote{More
  than 20 known Jupiter-sized transiting planets have a similar or
  higher density.}\\ Due to the specific dates of the RV observations,
there is a possibility that the RV variation is caused by a log-term
linear trend due to an undetected stellar companion of POTS-1 in which
case the true RV amplitude and the true mass of POTS-1b would be
lower. Also there could be a systematic offset between the
observations taken in the two different instrument configurations. As
a consequence, the true RV variation could be actually smaller or
larger than that estimated in our analysis and the true mass and
density of POTS-1b could be lower or higher. We therefore analysed the
RV data of the two instrumental configurations independently (see
Section \ref{subsec.spectroscopic3}) and found best planetary masses
of M$_{p}$\,=\,1.91$^{+1.30}_{-1.19}$ and
0.38$^{+0.86}_{-0.38}$\,M$_{\rm{Jup}}$, respectively, which are both
in the planetary regime.\\ In order to rule out a blend scenario, we
analysed the bisector span as a function of orbital phase which showed
no significant variations although the errors on the measurements are
rather large. In addition, we investigate the possibility to rule out
a blend scenario based on the light curves alone. We were able to set
an upper limit of $\sim$50 per cent on the light coming from a
blending source.\\ Using high-resolution imaging we identified a
contaminating source $\sim$1.3\,arcsec north-west of POTS-1; however,
the contamination in the $K_s$ band is as low as $\sim$2 per
cent. There is no indication for any additional bright blending
source.\\ Despite its faintness, the parameters of the POTS-1 system,
i.e. planetary radius and stellar density, inclination, orbital period
and epoch, are very well constraint thanks to the high-precision
observations with GROND and the long baseline between the WFI and
GROND data sets.\\ Among all transiting planets discovered so far,
POTS-1b is orbiting one of the coolest stars and the period of 3.16\,d
is one of the longest periods of all transiting planets detected
around mid/end-K dwarfs and M dwarfs.\\ In Section \ref{sec.POTS-C2},
we presented an unconfirmed candidate found in the POTS,
i.e. POTS-C2b, which has a period of
2.76303$\pm$2.0\,$\times$\,10$^{-4}$\,d. The host star has an
effective temperature of 4900$\pm$400\,K as derived from SED fitting,
and the best-fitting radius of the planet candidate is
1.30$\pm$0.21\,R$_{\rm{Jup}}$. Our follow-up observations of POTS-1
have shown that the confirmation of a planet orbiting a faint star can
be difficult and costly with currently available instrumentation. In
the optical wavelength regime, POTS-C2 is $\sim$0.6\,mag fainter than
POTS-1. Nevertheless, POTS-C2b may be an interesting target for
follow-up studies in the future.\\ As a final remark, we want to point
out that the main reason for not finding any candidates around
brighter stars is the limited survey area of this pilot project.\\
\section*{Acknowledgements}
We thank Robert Filgas and Marco Nardini for the GROND operation
during the observations.\\
Part of the funding for GROND (both hardware as well as personnel) was
generously granted from the Leibniz-Prize to Professor G. Hasinger
(DFG grant HA 1850/28-1).\\
This publication makes use of data products from the Two Micron All
Sky Survey, which is a joint project of the University of
Massachusetts and the Infrared Processing and Analysis
Center/California Institute of Technology, funded by the National
Aeronautics and Space Administration and the National Science
Foundation.\\
This research has made use of the NASA/IPAC Extragalactic Database
(NED) which is operated by the Jet Propulsion Laboratory, California
Institute of Technology, under contract with the National
Aeronautics and Space Administration.\\
Furthermore, we have made use of NASA's Astrophysics Data System as
well as the \mbox{SIMBAD} data base, operated at CDS, Strasbourg,
France.\\

\label{lastpage}


\begin{thebibliography}{85}

\bibitem[\protect\citeauthoryear{{Adelman-McCarthy}
    {et~al}\mbox{.}}{2007}]{2007ApJS..172..634A} {Adelman-McCarthy}
  J.~K. {et~al.}, 2007, ApJS, 172, 634

\bibitem[\protect\citeauthoryear{{Alard} \& {Lupton}}{1998}]{Alard}
  {Alard} C., {Lupton} R.~H., 1998, ApJ, 503, 325

\bibitem[\protect\citeauthoryear{{Am{\^o}res} \&
    {L{\'e}pine}}{2005}]{2005AJ....130..659A} {Am{\^o}res} E.~B.,
  {L{\'e}pine} J.~R.~D., 2005, AJ, 130, 659

\bibitem[\protect\citeauthoryear{{Asplund}
    {et~al}\mbox{.}}{2009}]{2009ARA&A..47..481A} {Asplund} M.,
  {Grevesse} N., {Sauval} A.~J., {Scott} P., 2009, ARA\&A, 47, 481

\bibitem[\protect\citeauthoryear{{Auvergne}
    {et~al}\mbox{.}}{2009}]{2009A&A...506..411A} {Auvergne} M. et~al.,
  2009, A\&A, 506, 411

\bibitem[\protect\citeauthoryear{{Baade}
    {et~al}\mbox{.}}{1999}]{1999Msngr..95...15B} {Baade} D. et~al.,
  1999, The Messenger, 95, 15

\bibitem[\protect\citeauthoryear{{Bakos}
    {et~al}\mbox{.}}{2004}]{2004PASP..116..266B} {Bakos} G., {Noyes}
  R.~W., {Kov{\'a}cs} G., {Stanek} K.~Z., {Sasselov} D.~D., {Domsa}
  I., 2004, PASP, 116, 266

\bibitem[\protect\citeauthoryear{{Bakos}
    {et~al}\mbox{.}}{2006}]{2006ApJ...641L..57B} {Bakos}, G.~{\'A}.,
  {P{\'a}l}, A., {Latham}, D.~W., {Noyes}, R.~W., {Stefanik}, R.~P.,
  2006, ApJ, 641, L57

\bibitem[\protect\citeauthoryear{{Baraffe}
    {et~al}\mbox{.}}{1998}]{1998A&A...337..403B} {Baraffe} I.,
  {Chabrier} G., {Allard} F., {Hauschildt} P.~H., 1998, A\&A, 337, 403

\bibitem[\protect\citeauthoryear{{Barbieri}}{2007}]{2007ASPC..366...78B}
  {Barbieri} M., 2007, in Afonso C., Weldrake D., Henning T., eds, ASP
  Conf. Ser., Vol. 366, Transiting Extrapolar Planets
  Workshop. Astron. Soc. Pac., San Francisco, p.~78

\bibitem[\protect\citeauthoryear{{Batalha}
    {et~al}\mbox{.}}{2012}]{2013ApJS..204...24B} {Batalha}
  N. M. et~al., 2013, ApJS, 204, 24

\bibitem[\protect\citeauthoryear{{Beatty} \&
    {Gaudi}}{2008}]{2008ApJ...686.1302B} {Beatty} T.~G., {Gaudi}
  B.~S., 2008, ApJ, 686, 1302

\bibitem[\protect\citeauthoryear{{Berta}
    {et~al}\mbox{.}}{2012}]{2012ApJ...747...35B} {Berta} Z. K. et~al.,
  2012, ApJ, 747, 35

\bibitem[\protect\citeauthoryear{{Bertin} \&
    {Arnouts}}{1996}]{1996A&AS..117..393B} {Bertin} E., {Arnouts} S.,
  1996, A\&AS, 117, 393

\bibitem[\protect\citeauthoryear{{Binney} \&
    {Merrifield}}{1998}]{1998gaas.book.....B} {Binney} J.,
  {Merrifield} M., 1998, {Galactic Astronomy. Princeton Univ. Press,
    Princeton, NJ}

\bibitem[\protect\citeauthoryear{{Borucki}
    {et~al}\mbox{.}}{2010}]{2010Sci...327..977B} {Borucki}
  W. J. et~al., 2010, Science, 327, 977

\bibitem[\protect\citeauthoryear{{Capaccioli}, {Mancini} \&
    {Sedmak}}{2002}]{2002SPIE.4836...43C} {Capaccioli} M., {Mancini}
  D., {Sedmak} G., 2002, in {Tyson} J.~A., {Wolff} S., eds, Proc. SPIE
  Conf. Ser. Vol. 4836, Survey and Other Telescope Technologies and
  Discoveries. SPIE, Bellingham, p. 43

\bibitem[\protect\citeauthoryear{Charbonneau
    et~al.}{2000}]{2000ApJ...529L..45C} Charbonneau D., Brown T.~M.,
  Latham D.~W., Mayor M., 2000, ApJ, 529, L45

\bibitem[\protect\citeauthoryear{Charbonneau
    et~al.}{2005}]{2005ApJ...626..523C} Charbonneau D. et al., 2005,
  ApJ, 626, 523

\bibitem[\protect\citeauthoryear{{Claret} \&
    {Bloemen}}{2011}]{2011A&A...529A..75C} {Claret} A., {Bloemen} S.,
  2011, A\&A, 529, A75

\bibitem[\protect\citeauthoryear{{Col{\'o}n}, {Ford} \&
    {Morehead}}{2012}]{2012MNRAS.426..342C} {Col{\'o}n} K.~D., {Ford}
  E.~B., {Morehead} R.~C., 2012, MNRAS, 426, 342

\bibitem[\protect\citeauthoryear{{de Mooij}
    {et~al}\mbox{.}}{2012}]{2012A&A...538A..46D} {de Mooij}
  E. J. W. et~al., 2012, A\&A, 538, A46

\bibitem[\protect\citeauthoryear{{Dekker}
    {et~al}\mbox{.}}{2000}]{2000SPIE.4008..534D} {Dekker} H.,
  {D'Odorico} S., {Kaufer} A., {Delabre} B., {Kotzlowski} H., 2000, in
  {Iye} M., {Moorwood} A.~F., eds, Proc. SPIE Conf. Ser. Vol. 4008,
  Optical and IR Telescope Instrumentation and Detectors. SPIE,
  Bellingham, p. 534

\bibitem[\protect\citeauthoryear{{Devillard}}{2001}]{2001ASPC..238..525D}
  {Devillard} N., 2001, in {Harnden} Jr.  F.~R., {Primini} F.~A.,
  {Payne} H.~E., eds, ASP Conf. Ser. Vol. 238, Astronomical Data
  Analysis Software and Systems X. Astron. Soc. Pac., San Francisco,
  p. 525

\bibitem[\protect\citeauthoryear{{Dotter}
    {et~al}\mbox{.}}{2008}]{2008ApJS..178...89D} {Dotter} A.,
  {Chaboyer} B., {Jevremovi{\'c}} D., {Kostov} V., {Baron} E.,
  {Ferguson} J.~W., 2008, ApJS, 178, 89

\bibitem[\protect\citeauthoryear{{Fossati}
    {et~al}\mbox{.}}{2009}]{2009A&A...503..945F} {Fossati} L.,
  {Ryabchikova} T., {Bagnulo} S., {Alecian} E., {Grunhut} J.,
  {Kochukhov} O., {Wade} G., 2009, A\&A, 503, 945

\bibitem[\protect\citeauthoryear{{Fressin}
    {et~al}\mbox{.}}{2007}]{2007A&A...475..729F} {Fressin} F.,
  {Guillot} T., {Morello} V., {Pont} F., 2007, A\&A, 475, 729

\bibitem[\protect\citeauthoryear{{Fukui}
    {et~al}\mbox{.}}{2011}]{2011PASJ...63..287F} {Fukui} A. et~al.,
  2011, PASJ, 63, 287

\bibitem[\protect\citeauthoryear{{G{\"o}ssl} \&
    {Riffeser}}{2002}]{2002A&A...381.1095G} {G{\"o}ssl} C.~A.,
  {Riffeser} A., 2002, A\&A, 381, 1095

\bibitem[\protect\citeauthoryear{{Greiner}
    {et~al}\mbox{.}}{2008}]{2008PASP..120..405G} {Greiner} J. et~al.,
  2008, PASP, 120, 405

\bibitem[\protect\citeauthoryear{{Gustafsson}
    {et~al}\mbox{.}}{2008}]{2008A&A...486..951G} {Gustafsson} B.,
  {Edvardsson} B., {Eriksson} K., {J{\o}rgensen} U.~G., {Nordlund}
  {\AA}., {Plez} B., 2008, A\&A, 486, 951

\bibitem[\protect\citeauthoryear{{Irwin}
    {et~al}\mbox{.}}{2009}]{2009IAUS..253...37I} {Irwin} J.,
  {Charbonneau} D., {Nutzman} P., {Falco} E., 2009, in Pont F.,
  Sasselov D., Holman M., eds, Proc. IAU Symp. Vol. 253, Transiting
  Planets. Cambridge Univ. Press, Cambridge, p. 37

\bibitem[\protect\citeauthoryear{{Jenkins}
    et~al.}{2010}]{2010ApJ...724.1108J}{Jenkins} J. M. et~al., 2010,
  ApJ, 724, 1108-1119

\bibitem[\protect\citeauthoryear{{Jester}
    {et~al}\mbox{.}}{2005}]{2005AJ....130..873J} {Jester} S. et~al.,
  2005, AJ, 130, 873

\bibitem[\protect\citeauthoryear{Koppenhoefer}{2009}]{2009PhDT.......287K}
  Koppenhoefer, J., 2009, PhD thesis, Ludwig Maximilians University
  Munich

\bibitem[\protect\citeauthoryear{{Koppenhoefer}
    {et~al}\mbox{.}}{2009}]{2009A&A...494..707K} {Koppenhoefer} J.,
  {Afonso} C., {Saglia} R.~P., {Henning} T., 2009, A\&A, 494, 707

\bibitem[\protect\citeauthoryear{{Koppenhoefer}, {Saglia} \&
    {Riffeser}}{2013}]{2013ExA....35..329K} {Koppenhoefer} J.,
  {Saglia} R.~P., {Riffeser} A., 2013, Exp. Astron., 35, 329

\bibitem[\protect\citeauthoryear{{Kov{\'a}cs}, {Zucker} \&
    {Mazeh}}{2002}]{2002A&A...391..369K} {Kov{\'a}cs} G., {Zucker} S.,
  {Mazeh} T., 2002, A\&A, 391, 369

\bibitem[\protect\citeauthoryear{{Kov{\'a}cs}
    {et~al}\mbox{.}}{2013}]{2013MNRAS.tmp.1446K} {Kov{\'a}cs}
  G. {et~al}\mbox{.}, 2013, MNRAS, 433, 889

\bibitem[\protect\citeauthoryear{{Kuijken}
    {et~al}\mbox{.}}{2002}]{2002Msngr.110...15K} {Kuijken} K. et~al.,
  2002, The Messenger, 110, 15

\bibitem[\protect\citeauthoryear{{Kupka}
    {et~al}\mbox{.}}{1999}]{1999A&AS..138..119K} {Kupka} F.,
  {Piskunov} N., {Ryabchikova} T.~A., {Stempels} H.~C., {Weiss} W.~W.,
  1999, A\&AS, 138, 119

\bibitem[\protect\citeauthoryear{{Kurucz}}{1993}]{1993KurCD..13.....K}
  {Kurucz} R., 1993, ATLAS9 Stellar Atmosphere Programs and 2 km/s
  grid.~Kurucz CD-ROM No.~13.~ Smithsonian Astrophysical Observatory,
  Cambridge, p. 13

\bibitem[\protect\citeauthoryear{{Landolt}}{1992}]{1992AJ....104..340L}
  {Landolt} A.~U., 1992, AJ, 104, 340

\bibitem[\protect\citeauthoryear{{Law}
    {et~al}\mbox{.}}{2011}]{2011ASPC..448.1367L} {Law} N.~M., {Kraus}
  A.~L., {Street} R.~R., {Lister} T., {Shporer} A., {Hillenbrand}
  L.~A., {Palomar Transient Factory Collaboration}, 2011, in
  {Johns-Krull} C., {Browning} M.~K., {West} A.~A., eds, ASP
  Conf. Ser. Vol. 448, 16th Cambridge Workshop on Cool Stars, Stellar
  Systems, and the Sun. Astron. Soc. Pac., San Francisco, p. 1367

\bibitem[\protect\citeauthoryear{{Lendl}
    {et~al}\mbox{.}}{2010}]{2010A&A...522A..29L} {Lendl} M., {Afonso}
  C., {Koppenhoefer} J., {Nikolov} N., {Henning} T., {Swain} M.,
  {Greiner} J., 2010, A\&A, 522, A29

\bibitem[\protect\citeauthoryear{{Lenzen}
    {et~al}\mbox{.}}{2003}]{2003SPIE.4841..944L} {Lenzen} R. et~al.,
  2003, in Iye M., {Moorwood} A.~F.~M., eds, Proc. SPIE
  Conf. Ser. Vol. 4841, Instrument Design and Performance for
  Optical/Infrared Ground-based Telescopes. SPIE, Bellingham, p. 944

\bibitem[\protect\citeauthoryear{{Maciejewski}
    {et~al}\mbox{.}}{2010}]{2010MNRAS.407.2625M} {Maciejewski}
  G. et~al., 2010, MNRAS, 407, 2625

\bibitem[\protect\citeauthoryear{{Mandel} \&
    {Agol}}{2002}]{2002ApJ...580L.171M} {Mandel} K., {Agol} E., 2002,
  ApJ, 580, L171

\bibitem[\protect\citeauthoryear{{McCullough}
    {et~al}\mbox{.}}{2005}]{2005PASP..117..783M} {McCullough} P.~R.,
  {Stys} J.~E., {Valenti} J.~A., {Fleming} S.~W., {Janes} K.~A.,
  {Heasley} J.~N., 2005, PASP, 117, 783

\bibitem[\protect\citeauthoryear{{Meeus}}{1982}]{1982QB51.3.E43M43..}
  {Meeus} J., 1982, {Astronomical Formulae for
    Calculators}. Willmann-Bell, Richmond, VA, p.~43

\bibitem[\protect\citeauthoryear{{Morel}, \&
    {Miglio}}{2012}]{2012MNRAS.419L..34M} {Morel}, T. and {Miglio},
  A., 2012, MNRAS, 419, 34

\bibitem[\protect\citeauthoryear{{Morton} \&
    {Johnson}}{2011}]{2011ApJ...738..170M} {Morton} T.~D., {Johnson}
  J.~A., 2011, ApJ, 738, 170

\bibitem[\protect\citeauthoryear{{Nikolov}
    {et~al}\mbox{.}}{2012}]{2012A&A...539A.159N} {Nikolov} N.,
  {Henning} T., {Koppenhoefer} J., {Lendl} M., {Maciejewski} G.,
  {Greiner} J., 2012, A\&A, 539, A159

\bibitem[\protect\citeauthoryear{{O'Donovan}
    {et~al}\mbox{.}}{2007}]{2007ApJ...663L..37O} {O'Donovan}
  F. T. et~al., 2007, ApJ, 663, L37

\bibitem[\protect\citeauthoryear{{Pasquini}
    {et~al}\mbox{.}}{2002}]{2002Msngr.110....1P} {Pasquini} L. et~al.,
  2002, The Messenger, 110, 1

\bibitem[\protect\citeauthoryear{{Pasquini}
    {et~al}\mbox{.}}{2012}]{2012A&A...545A.139P} {Pasquini} L. et~al.,
  2012, A\&A, 545, A139

\bibitem[\protect\citeauthoryear{{Pavlenko}}{1997}]{1997Ap&SS.253...43P}
  {Pavlenko} Y.~V., 1997, Ap\&SS, 253, 43

\bibitem[\protect\citeauthoryear{{Pavlenko}}{2003}]{2003ARep...47...59P}
  {Pavlenko} Y.~V., 2003, Astron. Rep., 47, 59

\bibitem[\protect\citeauthoryear{{Pavlenko}
    {et~al}\mbox{.}}{2012}]{2012MNRAS.422..542P} {Pavlenko} Y.~V.,
  {Jenkins} J.~S., {Jones} H.~R.~A., {Ivanyuk} O., {Pinfield} D.~J.,
  2012, MNRAS, 422, 542

\bibitem[\protect\citeauthoryear{{Piskunov}
    {et~al}\mbox{.}}{1995}]{1995A&AS..112..525P} {Piskunov} N.~E.,
  {Kupka} F., {Ryabchikova} T.~A., {Weiss} W.~W., {Jeffery} C.~S.,
  1995, A\&AS, 112, 525

\bibitem[\protect\citeauthoryear{{Pollacco}
    {et~al}\mbox{.}}{2006}]{2006PASP..118.1407P} {Pollacco}
  D. L. et~al., 2006, PASP, 118, 1407

\bibitem[\protect\citeauthoryear{{Pont}, {Zucker} \&
    {Queloz}}{2006}]{2006MNRAS.373..231P} {Pont} F., {Zucker} S.,
  {Queloz} D., 2006, MNRAS, 373, 231

\bibitem[\protect\citeauthoryear{{Queloz}
    {et~al}\mbox{.}}{2001}]{2001A&A...379..279Q} {Queloz} D. et al.,
  2001, A\&A, 379, 279

\bibitem[\protect\citeauthoryear{{Riffeser}, {Seitz} \&
    {Bender}}{2008}]{2008ApJ...684.1093R} {Riffeser} A., {Seitz} S.,
  {Bender} R., 2008, ApJ, 684, 1093

\bibitem[\protect\citeauthoryear{{Robin}
    {et~al}\mbox{.}}{2003}]{2003A&A...409..523R} {Robin}, A.~C.,
  {Reyl{\'e}}, C., {Derri{\`e}re}, S. and {Picaud}, S., 2003, A\&A,
  409, 523

\bibitem[\protect\citeauthoryear{{Rousset}
    {et~al}\mbox{.}}{2003}]{2003SPIE.4839..140R} {Rousset} G. et~al.,
  2003, in {Wizinowich} P.~L., {Bonaccini} D., eds, Proc. SPIE
  Conf. Ser. Vol. 4839, Adaptive Optical System Technologies II. SPIE,
  Bellingham, p. 140

\bibitem[\protect\citeauthoryear{{Ryabchikova}
    {et~al}\mbox{.}}{1999}]{1999PhST...83..162R} {Ryabchikova} T.~A.,
  {Piskunov} N.~E., {Stempels} H.~C., {Kupka} F., {Weiss} W.~W., 1999,
  Phys. Scr. T, 83, 162

\bibitem[\protect\citeauthoryear{Sahu
    {et~al}\mbox{.}}{2006}]{2006Natur.443..534S} Sahu
  K. C. {et~al}\mbox{.}, 2006, Nature, 443, 534-540

\bibitem[\protect\citeauthoryear{{Schlegel}, {Finkbeiner} \&
    {Davis}}{1998}]{1998ApJ...500..525S} {Schlegel} D.~J.,
  {Finkbeiner} D.~P., {Davis} M., 1998, ApJ, 500, 525

\bibitem[\protect\citeauthoryear{{Skrutskie}
    {et~al}\mbox{.}}{2006}]{2006AJ....131.1163S} {Skrutskie}
  M. F. et~al., 2006, AJ, 131, 1163

\bibitem[\protect\citeauthoryear{{Snellen}
    {et~al}\mbox{.}}{2009}]{2009A&A...497..545S} {Snellen}
  I. A. G. et~al., 2009, A\&A, 497, 545

\bibitem[\protect\citeauthoryear{{Snellen}
    {et~al}\mbox{.}}{2007}]{2007A&A...476.1357S} {Snellen} I.~A.~G.,
  {van der Burg} R.~F.~J., {de Hoon} M.~D.~J., {Vuijsje} F.~N., 2007,
  A\&A, 476, 1357

\bibitem[\protect\citeauthoryear{{Stetson}}{1987}]{1987PASP...99..191S}
  {Stetson} P.~B., 1987, PASP, 99, 191

\bibitem[\protect\citeauthoryear{{Stetson}}{2000}]{2000PASP..112..925S}
  {Stetson} P.~B., 2000, PASP, 112, 925

\bibitem[\protect\citeauthoryear{{Tamuz}, {Mazeh} \&
    {Zucker}}{2005}]{2005MNRAS.356.1466T} {Tamuz} O., {Mazeh} T.,
  {Zucker} S., 2005, MNRAS, 356, 1466

\bibitem[\protect\citeauthoryear{{Tomaney} \&
    {Crotts}}{1996}]{1996AJ....112.2872T} {Tomaney} A.~B., {Crotts}
  A.~P.~S., 1996, AJ, 112, 2872


\bibitem[\protect\citeauthoryear{{Torres}
    et~al.}{2011}]{2011ApJ...727...24T}{Torres G. et~al.}, 2011, ApJ,
  727, 24

\bibitem[\protect\citeauthoryear{{Tsymbal}}{1996}]{1996ASPC..108..198T}
  {Tsymbal} V., 1996, in {Adelman} S.~J., {Kupka} F., {Weiss} W.~W.,
  eds, ASP Conf. Ser. Vol. 108, M.A.S.S.: Model Atmospheres and
  Spectrum Synthesis. Atron. Soc. Pac., San Francisco, p. 198

\bibitem[\protect\citeauthoryear{Udalski
    et~al.}{2004}]{2004AcA....54..313U} {Udalski} A., {Szymanski}
  M.~K., {Kubiak} M., {Pietrzynski} G., {Soszynski} I., {Zebrun} K.,
  {Szewczyk} O., {Wyrzykowski} L., 2004, Acta Astron., 54, 313

\bibitem[\protect\citeauthoryear{{Udalski}
    {et~al}\mbox{.}}{2008}]{2008AcA....58...69U} {Udalski} A.,
  {Szymanski} M.~K., {Soszynski} I., {Poleski} R., 2008, Acta Astron.,
  58, 69

\bibitem[\protect\citeauthoryear{Valenti \&
    Fischer}{2005}]{2005ApJS..159..141V} {Valenti} J.~A., {Fischer}
  D.~A., 2005, ApJS, 159, 141

\bibitem[\protect\citeauthoryear{Valentijn
    et~al.}{2007}]{2007ASPC..376..491V} {Valentijn} E. A. et~al.,
  2007, in {Shaw} R.~A., {Hill} F., {Bell} D.~J., eds, ASP
  Conf. Ser. Vol. 376, Astronomical Data Analysis Software and Systems
  XVI. Astron. Soc. Pac., San Francisco, p. 491

\bibitem[\protect\citeauthoryear{Winn
    et~al.}{2009}]{2009ApJ...703L..99W} {Winn} J.~N., {Johnson} J.~A.,
  {Albrecht} S., {Howard} A.~W., {Marcy} G.~W., {Crossfield} I.~J.,
  {Holman} M.~J., 2009, ApJ, 703, L99

\bibitem[\protect\citeauthoryear{Young}{1967}]{1967AJ.....72..747Y}
  Young, A.~T., 1967, AJ, 72, 747

\end{thebibliography}
\end{document}